# Multi-scale and Multi-path Cascaded Convolutional Network for Semantic Segmentation of Colorectal Polyps

Malik Abdul Manan[a], Feng Jinchao[a], Muhammad Yaqub[a], Shahzad Ahmed[a], Syed Muhammad Ali Imran[b], Imran Shabir Chuhan[c], Haroon Ahmed Khan[d]

[a] *Beijing Key Laboratory of Computational Intelligence and Intelligent System, Faculty of Information Technology, Beijing University of Technology, Beijing 100124, China*
[b] *Department of computer science and information technology, Superior University, Lahore, Pakistan*
[c] *Interdisciplinary Research Institute, Faculty of Science, Beijing University of Technology, Beijing, China*
[d] *Department of Electrical and Computer Engineering, COMSATS University Islamabad (CUI), Islamabad, Pakistan*

**Abstract**

Colorectal polyps are structural abnormalities of the gastrointestinal tract that can potentially become cancerous in some cases. The study introduces a novel framework for colorectal polyp segmentation named the Multi-Scale and Multi-Path Cascaded Convolution Network (MMCC-Net), aimed at addressing the limitations of existing models, such as inadequate spatial dependence representation and the absence of multi-level feature integration during the decoding stage by integrating multi-scale and multi-path cascaded convolutional techniques and enhances feature aggregation through dual attention modules, skip connections, and a feature enhancer. MMCC-Net achieves superior performance in identifying polyp areas at the pixel level. The Proposed MMCC-Net was tested across six public datasets and compared against eight SOTA models to demonstrate its efficiency in polyp segmentation. The MMCC-Net's performance shows Dice scores with confidence interval ranging between $77.43 \pm 0.12$, (77.08, 77.56) and $94.45 \pm 0.12$, (94.19, 94.71) and Mean Intersection over Union (MIoU) scores with confidence interval ranging from $72.71 \pm 0.19$, (72.20, 73.00) to $90.16 \pm 0.16$, (89.69, 90.53) on the six databases. These results highlight the model's potential as a powerful tool for accurate and efficient polyp segmentation, contributing to early detection and prevention strategies in colorectal cancer.

*Keywords:* Colorectal Polyp, Semantic Segmentation, Cascaded Convolution Network, Feature Aggregation, Attention Modules

## 1. Introduction

Colorectal Cancer (CRC) has recently emerged as a prevalent, life-threatening disease that risks people's health, ranking as the third most common cancer worldwide and the second leading cause of mortality [1, 2]. Patients with CRC have a much lower chance of survival as the condition worsens. The majority of colorectal cancers, over 90%, originate from colorectal polyps [3]. The fatality rate of colorectal cancer globally is shown in Figure 1. The early detection and removal of colorectal polyps, the precursors to over 90% of CRC cases, are critical in reducing both the incidence and mortality rates of colorectal cancer. Recognizing and addressing these polyps promptly in clinical settings can play a pivotal role in decreasing the incidence and death rates of colorectal cancer. However, professionals' assessments of polyps in colonoscopy images are subjective and labor-intensive, potentially resulting in a missed rate of up to 25% [4], underscoring the urgent need for automated and reliable segmentation methods to localize polyps accurately, alleviating physicians' workload.

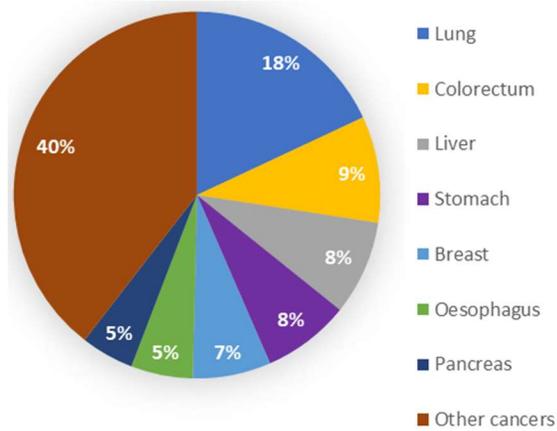

**Figure 1.** The number of estimated deaths because of cancer in 2020 worldwide, including Colorectum cancer in males and females of all ages.

Polyps are abnormal tissue growths that can appear on the body's surface, and the rectum, stomach, colon, and throat are places where they can be discovered [5]. Due to their potentially cancerous nature, accurate assessment is necessary in clinical settings, which includes analyzing changes in size, location, and possible malignancy. The polyps' various sizes and shapes pose serious difficulties during normal colonoscopies. Polyp lesion samples obtained under comparable circumstances may show problems with shape, size, background color, polyp count, and fuzzy boundaries, as shown in Figure 2. While examining medical images, it's challenging to distinguish these polyps as they often blend with surrounding tissue due to low contrast and unclear boundaries. It is essential to rapidly identify and remove CRC precursors, such as polyps, using accurate localization data to increase survival rates. Clinically, doctors can perform a colonoscopy to check the entire colon and remove polyps that have a propensity for CRC [6].

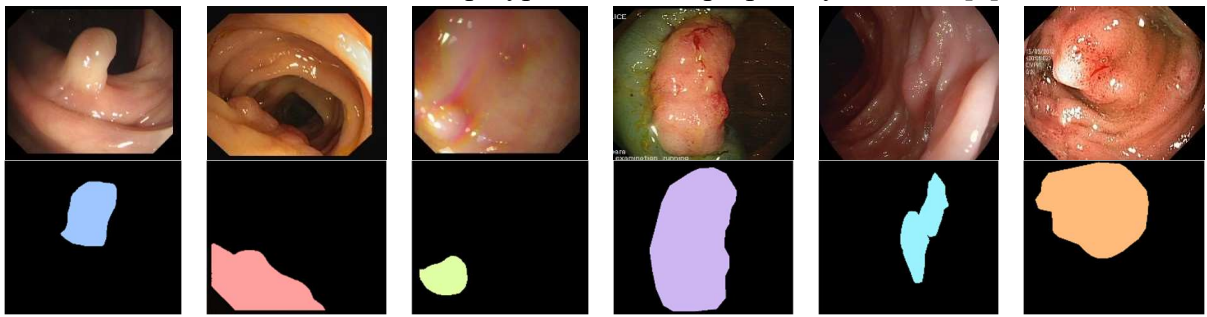

**Figure 2.** The image sample with a ground truth mask from all six databases we used in our research for segmentation.

The challenges in polyp segmentation are due to various factors, including the loss of local information, visual distractions due to polyp diversity, and the limitations of conventional segmentation algorithms that fail to provide a complete analysis. Deep learning (DL) for clinical diagnosis has revealed several challenges by linking polyp segmentation with conventional preliminary segmentation techniques. 1) Some DL model segmentation procedures use only global information from the high-level encoder, which may result in the loss of local information from lower-level layers. 2) Visual distractions due to differences in polyp shapes and sizes can make it difficult to segment borders and benign areas accurately. 3) Conventional segmentation algorithms frequently concentrate on particular elements of the diseased image and do not offer a whole analysis.

Our study provides a multi-scale cascading route network-based, effective, simplified, and powerful framework. The network's decoder utilizes cross-level features to produce high-quality predictions. Moreover, multi-level and multi-scale encoders account for various characteristics, including detailed representations and high-level details. Because polyp regions can vary in size, texture, and color, polyp region segmentation is typically thought to be difficult. Because polyp regions can be quite small, accurate detection calls for strong preprocessing, which drives up the cost of the system as a whole. The original colonoscopy image is input into the proposed multi-path, multi-scale cascaded convolution segmentation network (MMCC-Net), which then uses multiple deep feature aggregation coupled with multi-scale and multi-path convolutional processes. The powerful, comprehensive characteristics produced allow for precise pixel-wise findings of the polyp region.

## 1.1 Objectives and Contribution

This paper develops a novel approach through the Multi-Scale and Multi-Path Cascaded Convolutional Network (MMCC-Net) designed to advance the semantic segmentation of colorectal polyps. The following aims and objectives drive our work:

**Novelty:** The MMCC-Net integrates multi-scale and multi-path cascaded convolutional processes with advanced feature aggregation techniques by skip connection and attention modules designed to enhance the precision of polyp segmentation by overcoming the limitations of existing models.

**Aim:** This study aims to develop an efficient, simplified, and powerful model for the semantic segmentation of colorectal polyps that improves detection accuracy and reduces computational requirements.

**Objectives:** The objective is to determine the positioning and advantages of MMCC-Net relative to the state-of-the-art (SOTA) models. The model combines context features from various scales through dense skip connections, effectively addressing the feature degradation issue that presents many existing models. Integrating low-level feature information into the primary segmentation process, employing attention modules and a feature enhancer to enhance the segmentation of smaller polyps, is critical to improving detection accuracy. The joint function of dice loss and cross-entropy loss for pixel-wise classification is a strategy designed to mitigate the effects of class imbalance in the dataset. The study significantly reduces the number of parameters the MMCC-Net requires, ensuring the network's computational efficiency for training and execution. These objectives collectively aim to advance the field of medical image segmentation by providing a more efficient, accurate, and robust solution for the detection and segmentation of colorectal polyps.

**Contributions:** The proposed MMCC-Net makes the following important contributions:
- The MMCC-Net is a multi-scale model that consolidates context features from various scales through dense skip connections and addresses the feature degradation issue by using attention modules and a feature enhancer, which integrate low-level feature information into the primary segmentation process for improved results. This feature enhancement allows the network to segment smaller spots while reducing the network's depth precisely.
- For pixel-wise categorization, MMCC-Net uses a joint function of dice loss and cross-entropy loss, which helps to counteract the effects of class imbalance in the dataset. In comparison to other deep neural networks, MMCC-Net has significantly lower parameters, approximately 1434054, making it more computationally efficient for training and execution while still delivering competitive or even superior performance.
- Extensive experiments were performed on six publicly available datasets with visual results to measure the evaluation's performance compared to eight SOTA techniques. By incorporating these elements, MMCC-Net offers a versatile and efficient solution for polyp segmentation that can increase detection precision and lower the process's computing requirements.

The sections of this paper are as follows: We discuss the deep learning-based techniques in Section 2. Our suggested methodology is discussed in Section 3. In section 4, we perform experiments to contrast our approach and comparison with SOTA approaches. Section 5 discusses the findings, and section 6 concludes the study.

## 2. Literature Work

Current DL techniques perform better at polyp segmentation tasks than conventional methods. The decoder uses deconvolution for up-sampling, while the encoder captures high-dimensional features to provide segmentation outputs like the input image size. As DL has advanced [7-26], researchers have developed the U-Net network structure to remedy the drawbacks of Fully Convolutional Networks (FCN). U-Net and FCN have common encoder-decoder and skip connections [27, 28]. HarDNet-MSEG [29] incorporates the RFB Module and Dense Aggregation to delineate boundary and non-boundary regions. However, this approach has significantly missed detections within the segmentation outcomes. By fusing superficial detail information with deep semantic knowledge, skip connections help DNN to perform very well in challenges requiring semantic segmentation of medical images [30, 31]. The U-Net structure has been used as the foundation for many high-performing segmentation networks. DMU-Net [32] designed an inverted U and direct U dual-path network to extract contextual information and enrich edge information. This network uses an aggregation module, a pyramid attention module (PAM), and a margin refinement module (MRM). This approach provides remarkable segmentation results on the MTNS dataset and assists medical professionals in providing more accurate diagnoses of thyroid problems. During the segmentation process, an augmentation technique called meta-learning mixup (MLMix) [33] was combined with an approach called CAR.

## 2.1 Attention-based Polyp Segmentation

Attention-based operations are popular for segmentation tasks, particularly polyp segmentation [34]. A dual attention-gated U-Net [35] has focus blocks based on channel and spatial attention modules. They developed edge-reverse and object-reverse attention blocks to connect the encoder and the decoder. An encoder-decoder-based architecture was developed to create masks and edges in polyp images [36]. A boundary improvement module for boundary-aware attention blocks was designed to create the polyp mask using encoder features and partial decoder output [37]. The study [38] developed a boundary-guided attention module that includes feature enhancement and spatial attention in each spatial dimension, supervised by the ground truth. The reverse attention [39] was introduced by considering the connection between polyp borders and regions, which resulted in a clearer delineation of polyp boundary regions.

Progressively normalized self-attention is a concept developed by researchers [40] that dynamically adjusts the receptive field. Their network takes into account both temporal and spatial information. In most cases, the qualities of the features at various levels in the encoder are distinct. For more precise segmentation, authors have focused on tackling challenges associated with the sizes and shapes of polyp areas [41]. This shift in emphasis is part of an ongoing effort to improve segmentation accuracy. PraNet [42] utilizes a recursive backward attention module and a parallel partial decoder (PPD) component to obtain global feature maps. After combining the high-level features through a parallel connection, the next step is to compute a global feature map using a PPD. After this, a global feature map and a method known as reverse attention (RA) are employed to construct the links between boundaries and regions, which ultimately leads to precise polyp region segmentation.

UACANet [43] improves overall context information by collecting data regarding unclear local locales. Each bottom-up flow prediction module calculates a saliency feature map, which is then passed on to the subsequent prediction module. The background, foreground, and uncertain area feature weights are computed by each prediction module using previously predicted saliency feature maps. These feature weights are then aggregated to construct associations between focal boundaries and regions. ACSNet [44] compiles information on various features using a module that can adapt to its surroundings. The local features are first sent from an encoder to a decoder layer, and then, with the help of local feature weighting, the emphasis is brought to bear on more intricate areas of the image. In the final step, channel attention accumulates contextual data for polyp segmentation. TGANet [45] uses a text-oriented attention mechanism for feature extraction while training. The model can do extra feature representation learning and easily segment polyps of varying sizes by adding an auxiliary classification job. This task is designed to weight feature information based on polyp regions.

## 2.2 Transformer-based Polyp Segmentation

In medical image-based diseases [46], overcoming the limitations of traditional methods is crucial. These methods [47] often focus only on the local appearances of polyps or fail to incorporate a multi-level feature representation that captures spatial dependencies during the decoding process. ColonFormer network [48] introduces an innovative encoder-decoder structure, which captures long-range semantic relationships through its encoder. It utilizes a transformer-based, lightweight design for global semantic analysis across multiple scales, and its decoder is designed hierarchically to enhance feature representation by learning multi-level features. Additionally, it incorporates a refinement module equipped with a novel skip connection technique to improve the accuracy of polyp boundary segmentation within the global map. Despite the successes of convolutional neural network (CNN)-based segmentation methods, their inherent limitations in accurately delineating polyp contours and locations prompt the need for advanced solutions.

Polyp-PVT [49] and Att-PVT [50] introduced a hybrid model that combined CNN with a Pyramid Vision Transformer (PVT) to overcome the challenges of polyp segmentation. Att-PVT network employed multidimensional information extraction for enhanced feature mapping, cascaded context integration for semantic depth, and a multi-level feature fusion module for improved boundary definition. LAPFormer [51] harnesses the hierarchical Transformer encoder for global feature extraction with a CNN decoder for local appearance delineation. The LAPFormer's decoder has a feature fusion module, enabling effective multi-scale feature correlation by merging features across different scales. Integrating global and local feature extraction techniques sets new benchmarks in the accuracy of polyp segmentation.

FCB-Swin [52] integrates a SwinV2 Transformer-UNET and demonstrates superior performance on renowned datasets Kvasir-SEG and CVC-ClinicDB. The study highlights the issue of data leakage from video sequences in the CVC-ClinicDB dataset, suggesting a more rigorous approach to data partitioning to avoid this problem. The TransUNet [53] combined the strengths of Transformers and U-Net architectures. Transformers encode image patches derived from CNN feature maps for global context extraction. The U-Net decoder component enhances the precision of localization by upsampling encoded features with high-resolution CNN feature maps. This approach leverages Transformers for their potent encoding capabilities and U-Net for its ability to refine spatial details.

Fu-TransHNet [54] combined Transformer and CNN for global and local feature learning. The network combined the global-local feature fusion (GLFF) to merge same-scale features and the dense fusion of multi-scale features (DFM) to boost features. Fu-TransHNet uses multi-view cooperative learning to assign importance weights to each branch and module, concluding in a well-rounded decision-making process. The TransNetR [55] combined a pre-trained ResNet50 encoder with three decoder blocks and ended with an upsampling layer, forming an efficient encoder-decoder network. By leveraging the robust feature extraction capabilities of ResNet50 and the precise segmentation potential of the decoder layers, TransNetR enhanced diagnostic performance in identifying colon polyps, a significant advancement in medical imaging. Unlike previous methods, CASCADE [56] integrates convolutional layers into transformer architectures, effectively capturing both long-range dependencies and local pixel relationships and achieved through a unique combination of an attention gate for feature fusion with skip connections and a convolutional attention module designed to refine the focus on relevant areas by diminishing background noise. Furthermore, the method employs a multi-stage framework for feature aggregation, significantly enhancing segmentation performance.

### 3. Methodology

In the methodology, we explain our new MMCC-Net architecture in detail, outlining its structure and the main principles behind its design. One of the main reasons we developed MMCC-Net was to address the challenges of polyp segmentation. Polyps have various sizes, shapes, colors, and patterns, making them difficult to segment. Polyps often look very similar to the surrounding tissue, making it hard to tell them apart. With this in mind, the MMCC-Net has been carefully designed to handle these challenges, ensuring accurate segmentation even in difficult situations.

### 3.1. MMCC-Net Architecture

The MMCC-Net architecture, illustrated in Figure 3, is based on four fundamental design principles that collectively enable accurate polyp segmentation:

#### 3.1.1. Multi-Path and Multi-Scale Strategy

The MMCC-Net's design is integrating a multi-path and multi-scale strategy. The central part of Figure 3 shows multiple stacked layers (represented by color bars). These layers, numbered from 1 to 116, represent the feature extraction part of the network. The network captures features with each subsequent layer at various scales, providing a hierarchical understanding of the input image. 3 x 3 convolution represents convolution operation with a 3x3 filter. BN + ReLu represents a batch normalization (BN) followed by a ReLU activation function. This combination aids in stabilizing the learning process and introducing non-linearity. In blocks 44 and 25, dilated convolution was initially used. The advantage of the dilated convolution is that it allows a network to have a larger receptive field without increasing the number of parameters. It is usually adopted when capturing larger spatial contexts. Unlike blocks 62 and 42, the purpose of using block 23 is not to up-sample or increase the spatial resolution of feature maps. Instead, it handles features where the immediate spatial context is more relevant than a broader context and implicitly focuses on more localized feature extraction.

Average pooling is used instead of other pooling methods, such as max pooling, because it preserves the average features in local patches, thereby reducing spatial dimensions without discarding too much representational information. This results in smoother feature maps, mitigating the risk of overemphasizing extreme values, which could introduce artifacts. Additionally, average pooling has shown to be less prone to overfitting in some scenarios, as it provides a more generalized representation by considering all activations in the pool region rather than just the maximum ones. In the context of our MMCC-Net, average pooling demonstrated effective performance in our experiments, aligning well with our segmentation objectives.

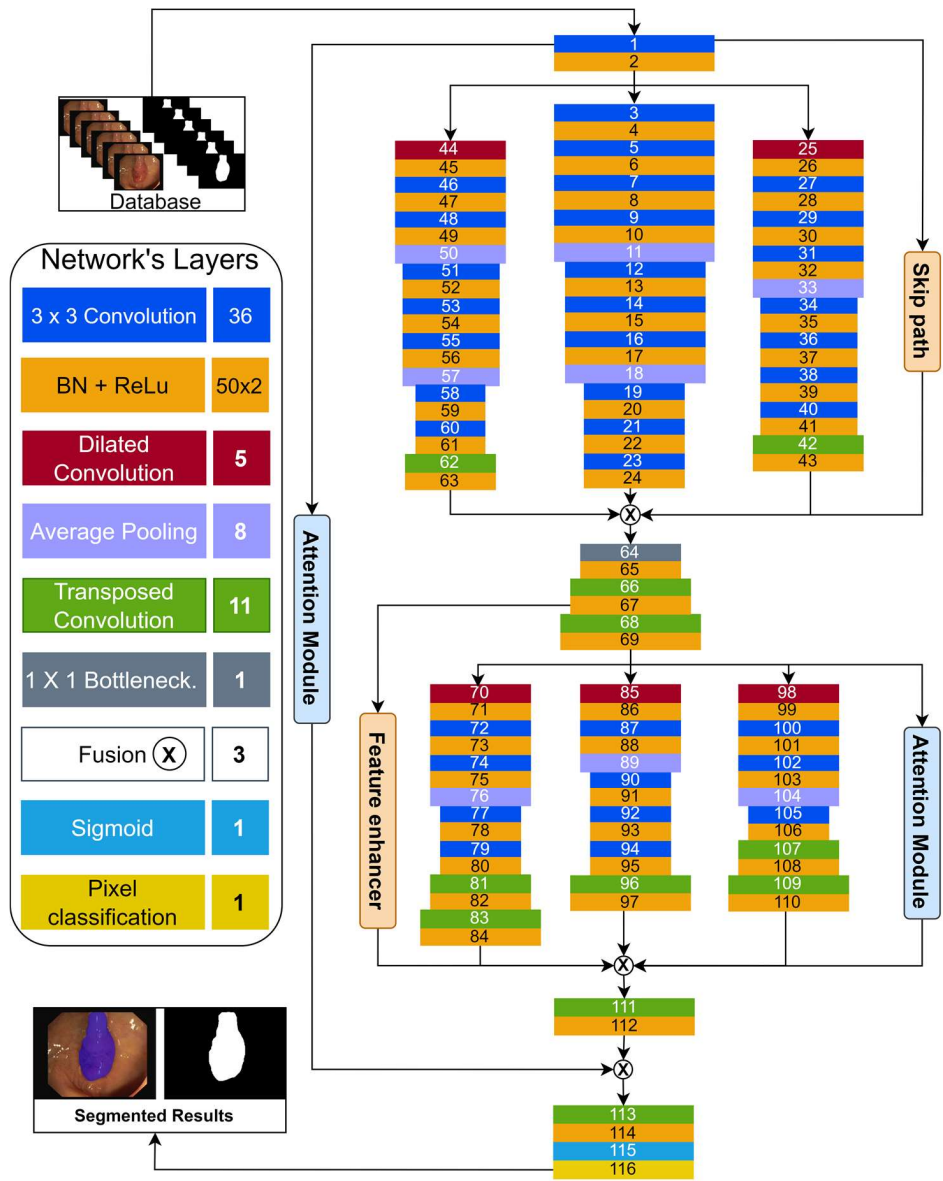

**Figure 3.** The architecture of the proposed MMCC-Net was tested on six publicly available datasets.

In our algorithm, spatial dimensions are reduced due to average pooling during the forward pass of many convolutional architectures. Moreover, these operations are beneficial for capturing hierarchical features. However, the size of the output image should match the size of the original input image. Considering this, transposed convolution is used to upscale these intermediate representations to a desired size. 1 x 1 bottleneck Reduces the number of channels to compress information and reduce computational cost. By introducing this strategy, MMCC-Net can harness the power of feature representations from multiple dimensions. The multi-path aspect ensures that features can be processed and captured through various channels or pathways. The multi-scale approach generates context features that completely understand the image, from finer details to broader structures captured at different resolutions. To enhance the effectiveness of this strategy, we've employed dense networking to aggregate these multi-scale features. Such a network topology is pivotal in mitigating potential feature latency issues. Traditional methods integrate features sequentially, often delaying processing. Dense networking addresses this by facilitating more direct connections and ensuring features are merged in a more integrated and timely manner.

### 3.1.2. Dense Skip Path and Attention Route for Edge Information

The MMCC-Net highlights the edges, which is crucial for accurately segmenting polyps. The model uses three different pathways designed to save edge information seamlessly. Dense skip and attention module routes carry valuable spatial information from the initial layers. Both components act as channels, ensuring that valuable spatial data from the early processing stages isn't lost but is effectively guided further into the network. The dense skip path provides a short route from earlier to later layers, helping the network to use low-level and high-level features in the segmentation process. By aggregating this spatial data, the MMCC-Net can differentiate even the most refined edge corners, strengthening its proficiency in precisely delineating polyps.

### 3.1.3. Attention Module and Feature Enhancer

The MMCC-Net has a strategic approach to ensure the segmentation of even the smallest polyps that can easily go unnoticed during convolution and pooling. The attention module and feature enhancer components are designed to refine the feature maps obtained from the convolution layers. The attention module focuses on significant parts of the polyp image, while the feature enhancer highlights important features. At its core, this attention module is a game-changer; the traditional pooling operations lead to the downscaling of feature maps, and our module maintains the original feature map size. It accomplishes this by utilizing a select set of convolutional layers, ensuring that despite having a more compact structure, no compromise is made on the efficacy of segmentation. The feature enhancer (FE) is pivotal, especially when segmenting diminutive features. The role of FE is important in the earlier layers of MMCC-Net, where FE captures essential low-level spatial cues like edge intricacies and maintains the integrity of the feature map's dimensions. The FE blends the responsiveness of ReLU with the normalization ability of BN. A series of four 3x3 convolutions then followed this merger. We've strategically limited the convolutional filters to 8 or 16 to ensure efficiency and minimize the computational overhead. This decision ensures that our model remains lightweight without sacrificing the fidelity of high-resolution features. Ultimately, this technique prevents the risks associated with pooling operations, which otherwise compromise spatial information integrity.

### 3.1.4. Parameter Efficiency

MMCC-Net carefully determines the number of filters used in each layer to ensure efficient model training and inference. Traditional networks often use excessive parameters, leading to overfitting and computational inefficiency. In contrast, the design of MMCC-Net achieves reliable polyp area segmentation while utilizing a modest number of trainable parameters—only 1434054 parameters, including 1428290 trainable and 5764 non-trainable parameters. This comprehensive architecture, combining multi-scale strategies, attention mechanisms, dense skip route, and parameter efficiency, equips MMCC-Net to address the challenges posed by diverse polyp characteristics and produce accurate polyp segmentation. After obtaining feature maps from the multi-path cascade network, the Sigmoid operation classifies each pixel for polyp segmentation and has demonstrated its effective performance.

Figure 4 explains the feature aggregation approach that was used so that one can have a better understanding of the connectivity pattern of MMCC-Net. MMCC-Net uses the input image as feature F and transforms it into $F_a$ after the first convolution; the features $F_a$ are sent to three different routes employing BN and ReLu, and the feature changes from $F_a$ to $F_i$. The features $F_i$ are sent into three different routes: a standard 3 by 3 convolution, which produces the feature $F_1$; a strided convolution with dilation and stride of 4 factors, which produces the feature $F_4$; a strided convolution with dilation and stride of 2 factors, which produces the feature $F_2$, and the attention module which produces the feature $AF_A$. When paired with the features from the skip path and three multi-scale features, the dense feature $DF_A$ is produced:

$$DF_A = F_a \otimes F_1 \otimes F_2 \otimes F_4 \tag{1}$$

Here, $\otimes$ signifies the depth-wise concatenation of dense features. This fusion approach allows us to combine feature maps from various layers, ensuring that spatial and contextual information from different levels is effectively integrated. The mid-block processes the attention module feature $AF_i$, and the feature $E_i$ is produced after the appropriate up-sampling. After that, dilated convolutions are used to build multi-scale features using this feature as the starting point. The first feature, denoted by $E_4$, results from the dilated convolution performed with a dilation and stride 4 factors. In contrast, the second feature, denoted by $E_2$, results from the dilated convolution performed with dilation and stride 2 factors, respectively. The dilated convolution with a dilation and stride 1 factor, which yields $E_1$, produces the third feature. Producing the dense feature, $DF_B$ is similar to the first stage except for the feature enhancer. Together with those from another dense skip path and the feature enhancer $AF_B$, these three features are accumulated to generate the $DF_B$.

$$DF_B = AF_i \otimes E_1 \otimes E_2 \otimes E_4 \otimes AF_B \tag{2}$$

To integrate minor edge information and spatial information from the first layers, the second dense component $DF_B$, which represents a powerful feature, is coupled after the transpose layer with the first $AF_A$ feature to generate a rich feature $DF_C$.

$$DF_C = D^l F_B \otimes AF_A \tag{3}$$

**3.2 Class Imbalance Problem**

The most widely used approach for implementing a weighted loss function involves assigning greater importance to the minority class while assigning lesser significance to the majority class. This can be achieved by setting the weights in inverse proportion to the class frequencies, ensuring that the minority class is given a higher weight and the majority class a lower weight. In the realm of medical image segmentation, class imbalance is a prevalent issue. This imbalance often leads the learning process to converge prematurely to local optima in the loss function, resulting in disproportionately biased predictions toward the background. To overcome this problem, we employ an $L_2$ loss on the Dice coefficient, eliminating the need to assign disparate weights to different classes to counterbalance the missed pixels from foreground and background. Figure 5 presents the Grad-CAM heatmaps generated by our suggested model, highlighting the critical features that play a role in determining the class. Grad-CAM visualizes the feature maps' average across the channels, where a high confidence level is represented by red, and blue signifies evidence of the class. It is observed that the MMCC-Net progressively learns to identify the polyp area fairly. Consequently, the loss formula we adopt is:

$$L = \sum_{I \, \varepsilon \, DB}(1 - D_{img})^2 \quad (4)$$

$D_{img}$ is the Dice coefficient for an image (img) that is stored in the database DB is written as:

$$D_{img} = \frac{2 \sum_{i \, \varepsilon \, img} p_i \widehat{p_i}}{\sum_{i \, \varepsilon \, img} p_i^2 \sum_{i \, \varepsilon \, img} \hat{p}_i^2} \quad (5)$$

The binary predicted label for pixel i in the image (img), which is a part of the dataset DB, is represented here by p$i$, and the matching ground truth is denoted by $\hat{p}_i$. These factors are taken into account to compute the gradient of the loss stated in Equation (5) as compared to the expected segmentation label:

$$\nabla\text{pi} \, L = 2(1 - D_{img}) \frac{\partial D_{img}}{\partial p} \quad (6)$$

By using the expression, the partial derivative for the dice coefficient can be calculated as:

$$\frac{\partial D_{img}}{\partial \text{pi}} = \frac{\hat{p}_i(\sum_{i \, \varepsilon \, img} p_i^2 + \sum_{i \, \varepsilon \, img} \hat{p}_i^2) - 2\text{pi}(\sum_{i \, \varepsilon \, img} p_i \widehat{p_i})}{(\sum_{i \, \varepsilon \, img} p_i^2 + \sum_{i \, \varepsilon \, img} \hat{p}_i^2)^2} \quad (7)$$

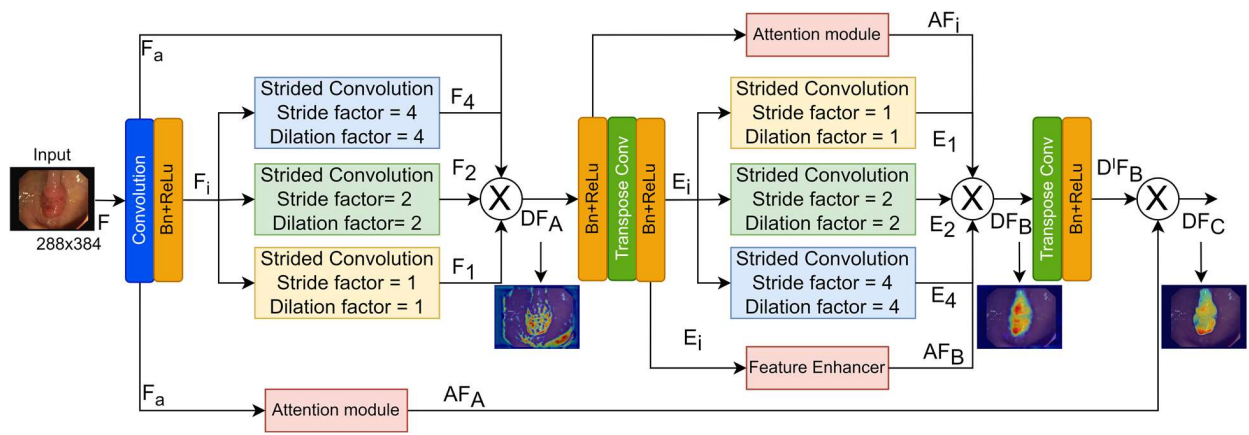

**Figure 4.** The dense features from multi-scale and multi-path by the proposed MMCC-Net.

## 3.3 Joint Loss Function

In our method, we combine the strengths of Dice loss and Binary Cross Entropy (BCE) loss to finely tune the segmentation of polyps. BCE loss prioritizes challenging sample pixels by assigning them more significance, whereas Dice loss offers meticulous pixel-level guidance. As we refine the MMCC-Net, profound supervision is employed, and the resulting segmentation loss can be expressed as:

$$L_{seg} = L_{Dice} + L_{Bce} \tag{8}$$

$$L_{Dice} = \left(1 - \frac{\sum_j^N y_i x_j}{\sum_j^N y_j^2 + \sum_j^N x_j^2 - y_j x_j}\right) \tag{9}$$

$$L_{Bce} = -\sum_j^N \left(y_j \log x_j + (1 - y_j) \log(1 - x_j)\right) \tag{10}$$

$$L_{seg} = \alpha \left(1 - \frac{\sum_j^N y_i x_j (-\alpha(1-x_j)^\gamma)}{\sum_j^N y_j^2 + \sum_j^N x_j^2 - y_j x_j}\right) + \left(-\sum_j^N \left(y_j \log x_j + (1 - \alpha) x_j^\gamma (1 - y_j) \log(1 - x_j)\right)\right) \tag{11}$$

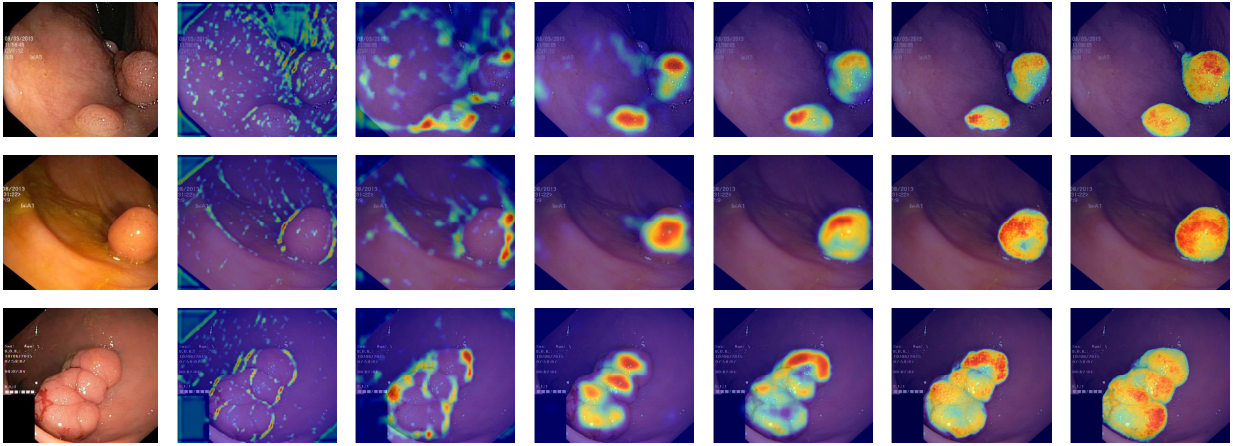

**Figure 5.** Grad-CAM heat map visualizations for MMCC-Net stages. Column 1 shows input images; the initial convolutional block heat maps are presented in Column 2; Columns 3-5 show results after the first, second, and third concatenation; Columns 6-7 present the results with dice loss and joint loss with sigmoid function respectively.

$L_{Bce}$ symbolizes the binary cross-entropy loss, focusing on individual pixel-level adjustments, while $L_{Dice}$, which addresses broader segmentation constraints, stands for the Dice loss. In this context, $L_{seg}$ denotes the loss function for the $i_{th}$ instance, where j indicates the pixel under consideration. The variable N encapsulates the entirety of pixels within the image and refers to the model's forecast for the pixel value at j. The coefficients α and γ serve as adjustable hyperparameters, facilitating the balance between the contribution of the Dice coefficient and the binary cross-entropy within the composite loss function. The settings for these parameters were 0.22 for α and 1.9 for γ in our experiments.

## 4. Experimental Setup and Results

### 4.1 Databases

We use six public datasets for comparison: EndoCV2020 [57], CVC-300 [58], ETIS [59], CVC-ColonDB [60], CVC-ClinicDB [61], and Kvasir [62]. These datasets are broadly used in evaluating the performance and efficacy of the various polyp segmentation methods under consideration. Extensive tests are carried out on these six benchmark colonoscopy datasets, and thorough explanations of each dataset are provided to improve comprehension of their characteristics.

**Kvasir:** The collection originates from the Vestre Viken Health Trust in Norway and comprises 1,000 images of polyps. These images come in various formats and sizes and are captured during gastrointestinal tract checkups.

**CVC-ClinicDB:** The dataset includes 612 images from 29 separate colonoscopy videos. The images have a size of 288x384 pixels. The visual richness and diversity of the CVC-ClinicDB dataset help test the effectiveness and efficiency of polyp segmentation methods.

**CVC-ColonDB:** The dataset contains 380 images with size of 500x570 pixels. The large image collection illustrates a wide range of colon conditions. The great quality of the images allows for a close examination of the colon's characteristics.

**ETIS:** There are 196 polyp images in the dataset. The image size is 966x1225 pixels and shows the intricate details of the polyps. This important dimension helps researchers and machine learning systems find and recognize small polyps.

**CVC-300:** The dataset contains 60 polyp images, an important contribution by the Computer Vision Center. The image size is 500x574 pixels, which provides fine detail for analyzing the polyps.

• EndoCV2020: EndoCV2020 is made up of 122 polyp images. This collection contains images with resolutions ranging from 1232x1048, 1463x1065, 572x498, 1350x1080, 628x513, and 392x365. Various resolution sizes allow for greater detail, and polyp features to be noted in the images.

### 4.2 Evaluation Measures

The study utilizes eight well-known quantitative evaluation measures: Dice, accuracy, sensitivity, precision, specificity [3], IOU, AUC, confidence interval (CI), and Hausdorff Distance (HDD) [63]. Precision is the ratio of the correctly predicted true positives to the total number of positives. Sensitivity (SN) or recall [64] measures the percentage of actual positive cases that are accurately segmented. Dice is a similarity measure function that determines the degree of similarity between the predicted outcome and the mask, calculated by taking the harmonic mean of precision and recall. Jaccard represents the ratio of the intersection among the predicted masks and the mask images to their union. AUC is the area under the ROC curve, which summarizes the model's overall performance. The Hausdorff Distance (HDD) quantifies the disparity between two sets of points, with a lower value indicating a closer resemblance between the pair of samples. The

mathematical expressions for each of these performance measures are:

$$\text{Accuracy} = \frac{(TP + TN)}{(TP + TN + FP + FN)} \tag{12}$$

$$\text{Precision} = \frac{TP}{(TP + FP)} \tag{13}$$

$$\text{Recall or sensitivity} = \frac{TP}{(TP + FN)} \tag{14}$$

$$\text{Dice} = \frac{TP}{(2TP + FN + FP)} \tag{15}$$

$$\text{IOU or Jaccard} = \frac{TP}{(FP + T + FN)} \tag{16}$$

$$\text{HDD} = \max\{max_{a \in A} min_{b \in B} \|a - b\|, max_{a \in A} min_{b \in B} \|b, a\|\} \tag{17}$$

$$\text{Confidence Interval} = \bar{x} \pm \text{Margin of Error} \tag{18}$$

Here, TP is true to classify polyp pixels, FP is false to classify non-polyp pixels, TN is true to classify non-polyp pixels, and FN is classified incorrectly as non-polyp pixels. 'A' refers to the actual image, while 'B' denotes the segmented result. The notation $\|a - b\|$ signifies the method used to calculate the distance between the two. $\bar{x}$ is the mean value of performance measures, and the margin of error is the product of the critical value tα/2t and the standard deviation divided by the square root of the sample size. In this case, the sample size is 10.

### 4.3 Result Analysis and Comparison with the SOTA Model

The model was developed and trained using MATLAB version 2023b. We initially ran our method for 80 epochs, which took approximately 8 hours. However, we observed no significant improvement in Metrics such as training loss, validation loss, training accuracy, and validation accuracy observed when the epoch was beyond 60, which took approximately 6 hours. During training, a separate validation set was used to monitor the model's generalization performance. Early stopping was implemented to stop training if the validation performance did not improve for a predefined number of epochs. Our approach is evaluated against eight SOTA models across six benchmark datasets, such as UNet [65], PraNet [42], HarDNet-MSEG [29], ColonFormer [48], FCB Former [47], Polyp-PVT [49], FCB-SwinV2 [52], and PVT-CASCADE [56], to illustrate its effectiveness and advancements. The open-source code for each of the eight models was employed for evaluation, utilizing identical training and test sets, with no optimization strategies applied. The original images were down-sampled, reducing their resolution to a consistent size of 288 x 384 pixels, the same size as the image collection of the CVC-ClinicDB dataset. This size reduction not only standardizes the input for our model but also makes the computational process more efficient without significantly sacrificing the quality or the essential details present in the images. This enables the model to handle and analyze large data more efficiently while maintaining the vital characteristics necessary for accurate polyp segmentation. A thorough analysis of the results will also shed light on key findings, challenges faced, and the implications for future research and practical applications.

Table 1. An overview of how the datasets are divided and used for training and testing in this study.

| Sr. No | Dataset | Total Images | Train images | Validation Images | Test images |
|---|---|---|---|---|---|
| 1 | Kvasir | 1000 | 800 | 100 | 100 |
| 2 | CVC-ClinicDB | 612 | 490 | 61 | 61 |
| 3 | CVC-300 | 60 | 0 | 0 | 60 |
| 4 | ETIS | 196 | 0 | 0 | 196 |
| 5 | CVC-ColonDB | 380 | 0 | 0 | 380 |
| 6 | EndoCv2020 | 122 | 0 | 0 | 122 |

### 4.3.1 Experimental setup

Our method is assessed through two experimental setups, detailed as follows:

**Experiment 1:** This experiment follows the partitioning guidelines recommended in previous studies [42, 48], allocating 90% of the data from the CVC-ClinicDB and Kvasir datasets to evaluate the model's learning capacity, including a 10% validation set. For testing, we use the remainder of the images from the Kvasir and CVC-ClinicDB datasets and all from the CVC-ColonDB, EndoCv2020, CVC-300, and ETIS-Larib datasets. The distribution of these datasets is precisely indicated in Table 1.

**Experiment 2:** We conduct a 5-fold cross-validation on the CVC-ClinicDB and Kvasir datasets, dividing each dataset into five equal parts [42, 48, 66]. Each iteration selects one part for testing and the four remaining parts for training. The Adam optimizer, with a learning rate of 1e-4, supports the training of MMCC-Net over 60 epochs and has a batch size of 8. MMCC-Net undergoes training five separate times, with the outcomes averaged across these instances for robustness.

**Experiment 3:** 10 independent runs (training and testing) were conducted for each model on the Kvasir and ClinicDB datasets. After training, the testing was performed on the remaining four datasets to ensure the robustness and reliability of our algorithm. The corresponding statistical results were summarized in Tables 2 and 4, including the mean values (mean) of five performance metrics, standard deviations (SD), and 95% confidence intervals (CI) for both the Kvasir and ClinicDB datasets and the mean value of three performance metrics, SD and 95% CI on remaining four datasets. These statistical evaluations demonstrate the consistent performance of our proposed method compared to other models.

**Results:** The evaluation of the model's learning capabilities is visually presented in Figures 6-9. In these figures, green represents the ground truth (GT) (true positives), yellow shows accurately predicted polyps, and red highlights incorrect predictions. The performance metrics are detailed in Tables 2-4, highlighting the top-performing.

Figure 6 illustrates a side-by-side comparison of our model against MMCC-Net, focusing on the Kvasir and CVC-ClinicDB datasets. The comparison shows the original images and their respective GT in the first two columns. The third column displays the results of the UNet model, highlighting polyps' diverse shapes and sizes and their color and texture similarities with adjacent tissues that significantly complicate precise polyp segmentation. Despite the U-Net [65] model's efforts to bridge different levels of feature information through skip connections, it struggles with capturing long-range feature dependencies, often missing smaller polyps in the segmentation process. Models like PraNet [42] and HarDNet-MSEG [29], lacking in Transformer technology for global context understanding, show considerable red areas, indicating mis-segmentations. Although HarDNet-MSEG recently introduced a multi-branch feature interaction approach, the absence of a mechanism to focus on global context information leads to segmentation inaccuracies and overlooked areas.

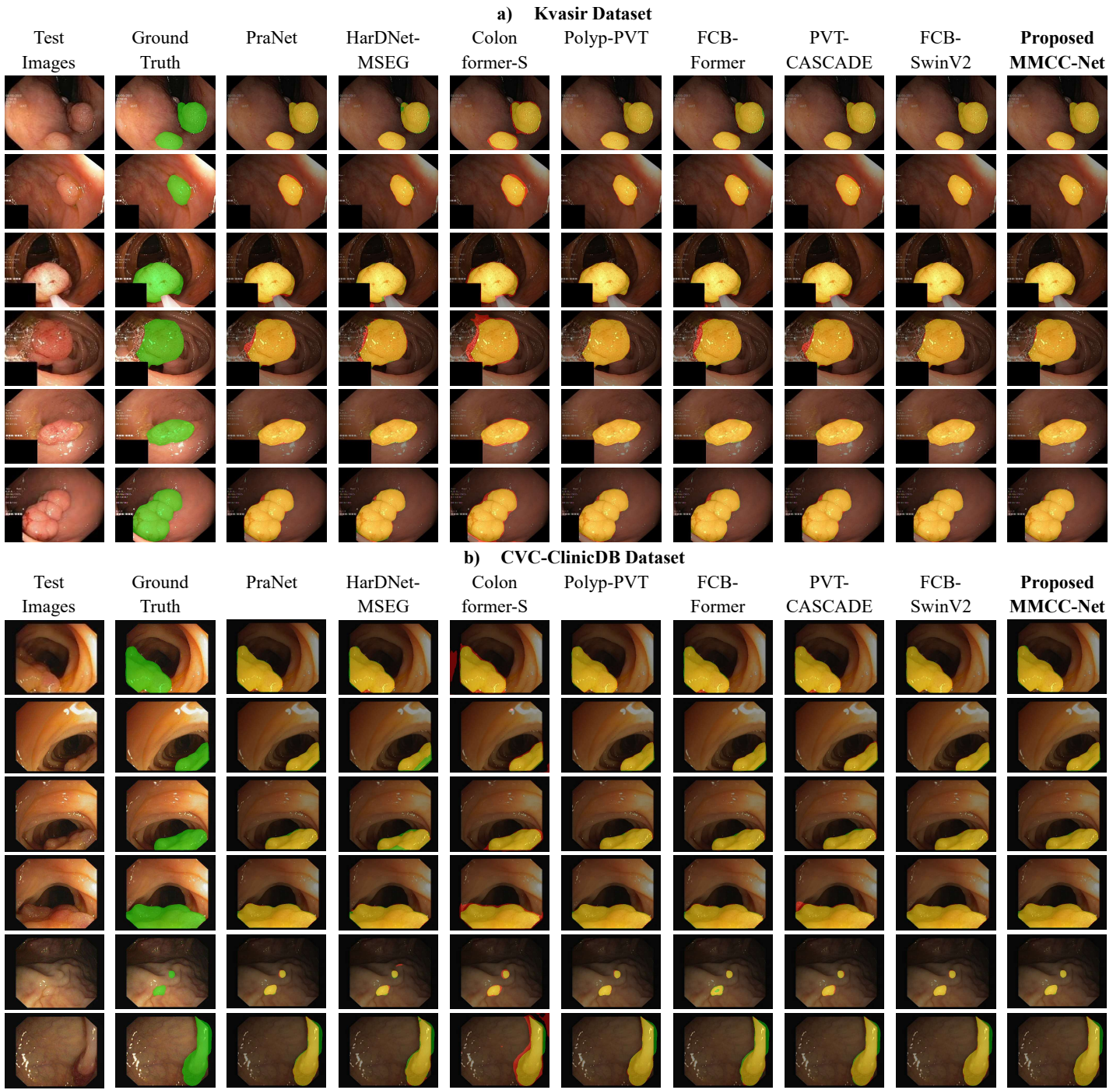

**Figure 6.** The result of our proposed MMCC-Net and Seven SOTA Models on the Kvasir and CVC-ClinicDB datasets. Columns 1-2 show the original images and Ground Truth of the datasets; columns 3-9 show the segmented results of SOTA, and column 10 shows the proposed MMCC-net.

**Table 2.** The comparisons of our proposed MMCC-net on two publicly available datasets with SOTA models for polyp segmentation with mean performance measures, standard deviation, and confidence interval (10 runs).

| Dataset | Method | Accuracy (Mean ± SD, 95% CI) | Dice (Mean ± SD, 95% CI) | MIoU (Mean ± SD, 95% CI) | Precision (Mean ± SD, 95% CI) | Recall (Mean ± SD, 95% CI) |
|---|---|---|---|---|---|---|
| **Kvasir [62]** | U-Net [65] | 83.34 ± 0.75, (82.68, 85.74) | 76.88 ± 0.72, (75.50, 78.46) | 73.40 ± 0.83, (71.70, 75.08) | 76.12 ± 0.78, (74.39, 77.57) | 78.35 ± 0.70, (76.16, 79.08) |
| | PraNet [42] | 94.26 ± 0.18, (93.83, 94.67) | 90.40 ± 0.23, (90.02, 91.04) | 85.81 ± 0.30, (85.12, 86.40) | 89.75 ± 0.27, (89.26, 90.36) | 92.22 ± 0.21, (91.70, 92.58) |
| | Polyp-PVT [49] | 96.18 ± 0.15, (95.91, 96.61) | 91.12 ± 0.19, (90.63, 91.51) | 86.35 ± 0.25, (85.91, 86.91) | 90.88 ± 0.21, (90.51, 91.41) | 93.21 ± 0.15, (92.81, 93.43) |
| | HarDNet-MSEG [29] | 95.31 ± 0.16, (94.85, 95.57) | 92.09 ± 0.20, (91.68, 92.56) | 85.68 ± 0.27, (85.17, 86.25) | 92.68 ± 0.18, (92.34, 93.14) | 91.51 ± 0.16, (90.98, 91.70) |
| | ColonFormer-S [48] | 95.43 ± 0.13, (95.25, 95.87) | 92.31 ± 0.17, (91.84, 92.68) | 87.72 ± 0.23, (87.26, 88.22) | 92.21 ± 0.15, (91.88, 92.54) | 92.95 ± 0.14, (92.66, 93.30) |
| | PVT-CASCADE [56] | 96.84 ± 0.13, (96.54, 97.06) | 92.35 ± 0.15, (92.03, 92.67) | 88.01 ± 0.19, (87.57, 88.45) | 94.32 ± 0.14, (94.05, 94.59) | 92.26 ± 0.13, (91.98, 92.54) |
| | FCB-Former [47] | 96.82 ± 0.11, (96.52, 97.04) | 92.32 ± 0.14, (91.96, 92.60) | 87.68 ± 0.19, (87.32, 88.20) | 94.16 ± 0.12, (93.92, 94.46) | 92.44 ± 0.11, (91.96, 92.52) |
| | FCB-SwinV2 [52] | 96.86 ± 0.12, (96.59, 97.11) | 92.52 ± 0.12, (92.22, 92.82) | 88.19 ± 0.17, (87.77, 88.61) | 94.35 ± 0.11, (94.10, 94.60) | 92.36 ± 0.10, (92.10, 92.62) |
| | **Proposed MMCC-Net** | **96.94 ± 0.11, (96.63, 97.15)** | **92.65 ± 0.13, (92.65, 93.25)** | **88.35 ± 0.18, (87.86, 88.70)** | **94.96 ± 0.11, (94.61, 95.11)** | **92.94 ± 0.14, (92.65, 93.21)** |
| **ClinicDB [61]** | U-Net [65] | 82.69 ± 0.75, (81.43, 84.49) | 73.71 ± 0.72, (72.14, 75.10) | 66.48 ± 0.83, (63.81, 67.19) | 74.38 ± 0.78, (73.65, 77.83) | 72.41 ± 0.70, (70.86, 73.78) |
| | PraNet [42] | 95.33 ± 0.18, (94.86, 95.70) | 91.82 ± 0.23, (91.30, 92.32) | 88.52 ± 0.30, (87.84, 89.12) | 94.21 ± 0.27, (93.52, 94.62) | 91.88 ± 0.21, (91.28, 92.16) |
| | Polyp-PVT [49] | 96.24 ± 0.15, (95.86, 96.56) | 93.69 ± 0.19, (93.34, 94.22) | 89.98 ± 0.25, (89.43, 90.43) | 92.72 ± 0.21, (92.22, 93.12) | 94.31 ± 0.15, (93.91, 94.53) |
| | HarDNet-MSEG [29] | 95.71 ± 0.16, (95.29, 96.01) | 93.29 ± 0.20, (92.78, 93.66) | 88.34 ± 0.27, (87.72, 88.80) | 92.69 ± 0.18, (92.24, 93.04) | 93.47 ± 0.16, (93.00, 93.72) |
| | ColonFormer-S [48] | 96.81 ± 0.11, (96.52, 97.04) | 93.35 ± 0.14, (92.92, 93.56) | 89.52 ± 0.19, (88.99, 89.87) | 94.78 ± 0.12, (94.55, 95.09) | 93.22 ± 0.14, (92.87, 93.43) |
| | PVT-CASCADE [56] | 97.02 ± 0.12, (96.78, 97.26) | 94.41 ± 0.14, (94.15, 94.67) | 90.01 ± 0.17, (89.59, 90.43) | 94.33 ± 0.14, (94.11, 94.55) | 95.61 ± 0.15, (95.41, 95.81) |
| | FCB-Former [47] | 96.92 ± 0.13, (96.74, 97.22) | 94.39 ± 0.13, (94.08, 94.60) | 89.97 ± 0.18, (89.56, 90.40) | 94.28 ± 0.13, (94.03, 94.47) | 95.52 ± 0.13, (95.28, 95.68) |
| | FCB-SwinV2 [52] | 97.10 ± 0.11, (96.86, 97.34) | 94.43 ± 0.13, (94.17, 94.69) | 90.08 ± 0.18, (89.66, 90.50) | 94.42 ± 0.12, (94.20, 94.64) | 95.67 ± 0.12, (95.47, 95.87) |
| | **Proposed MMCC-Net** | **97.29 ± 0.10, (96.95, 97.43)** | **94.45 ± 0.12, (94.19, 94.71)** | **90.16 ± 0.16, (89.69, 90.53)** | **94.44 ± 0.13, (94.22, 94.66)** | **96.81 ± 0.14, (96.63, 97.03)** |

The proposed MMCC-Net leverages a cascading CNN architecture to extract local spatial details and fine-grained information, significantly enhancing polyp localization precision while minimizing identification inaccuracies. Furthermore, integrating attention mechanisms and feature enhancer blocks allows for comprehensive global context comprehension, elevating the model's scene analysis capabilities and segmentation reliability. A pivotal aspect of our method is the innovative three-level feature fusion, wherein each tier is further branched into three distinct pathways to manage respective features, facilitating effectiveness across various scales to address the challenges posed by polyps' multi-scale variations. Visual inspection reveals that our model exhibits exceptional proficiency in segmenting polyps, particularly those of smaller size or with coloration that blends into the background and larger polyps delineated with complete shapes and distinct boundaries. Figure 6 (b), the visual comparison across methods on the CVC-ClinicDB dataset, highlights the original image in the first column, followed by the ground truth (GT) in the second. The ninth column shows the results from our proposed method. While PraNet [42] utilizes an attention mechanism to refine segmentation borders, it overlooks the semantic interplay between the area and its perimeter, leading to boundary-related segmentation inaccuracies and omissions.

The HarDNet-MSEG [29] model incorporates both the RFB Module and Dense Aggregation for delineating boundary and non-boundary regions and encounters significant shortcomings with missed detections within its segmentation outputs. Our methodology enriches the model with skip paths, attention mechanisms, feature enhancement modules, and depth-wise concatenation, employing dilated convolution for precise boundary delineation in segmentation tasks. The results reveal that segmentation boundaries consistently appear yellow, indicating flawless boundary accuracy without any segmentation errors. Conversely, models like ColonFormer [48], Polyp-PVT [49], and PVT-CASCADE [56] overlook the integration of varied level features and global contextual insights, leading to inaccuracies in the form of false positives and negatives in detection. ColonFormer's [48] structure captures long-range semantic relationships through its encoder, which utilizes a transformer-based, lightweight design for global semantic analysis across multiple scales, and its decoder, designed hierarchically to enhance feature representation by learning multi-level features. Polyp-PVT [49] introduced a hybrid model that synergizes CNN with a Pyramid Vision Transformer (PVT) to overcome these challenges. CASCADE [56] integrates convolutional layers into transformer architectures, often resulting in inconsistent feature representation. The method employs a multi-stage framework for feature aggregation, significantly enhancing segmentation performance. FCB-Former [47] and FCB-SwinV2 [52] used the power of transformer parallel with CNN shows superior results to other SOTA model but our approach simultaneously addresses the target and background areas, specifically targeting polyps' varying scales and similar textures. We significantly improve the network's scene comprehension by implementing multi-scale interaction and cascading fusion techniques. The visual outcomes demonstrate that our model precisely captures boundary information, resulting in minimal instances of incorrect segmentation.

The measures of 10 independent runs for each model on the Kvasir and ClinicDB datasets to validate the robustness and reliability of our proposed MMCC-Net were given in Table 2. The U-Net model showed moderate performance on both datasets. However, our proposed MMCC-Net demonstrated superior performance, consistently outperforming the SOTA models on both datasets. For the Kvasir dataset, MMCC-Net's accuracy improved by 16.34%, Dice score by 20.45%, and MIoU by 20.31% compared to U-Net. Compared to PraNet, MMCC-Net showed an improvement of 2.84% in accuracy, 2.49% in Dice score, and 2.96% in MIoU. When compared to Polyp-PVT, MMCC-Net's accuracy increased by 0.79%, Dice score by 1.68%, and MIoU by 2.31%. MMCC-Net outperformed HarDNet-MSEG with an improvement of 1.63% in accuracy, 0.56% in Dice score, and 2.67% in MIoU. Compared to ColonFormer-S, MMCC-Net's accuracy was higher by 1.51%, Dice score by 0.34%, and MIoU by 0.63%. MMCC-Net achieved a 0.10% higher accuracy, 0.30% higher Dice score, and 0.34% higher MIoU compared to PVT-CASCADE. The improvements over FCB-Former were 0.12% in accuracy, 0.33% in Dice score, and 0.67% in MIoU, and over FCB-SwinV2, the increases were 0.08% in accuracy, 0.13% in Dice score, and 0.16% in MIoU.

**Table 3.** Five-fold cross-validation comparing MMCC-Net and other models on the Kvasir and CVC-ClinicDB datasets are shown, with all outcomes expressed as the average value plus or minus the standard deviation across the five iterations.

| Dataset | Method | Dice | MIoU | Precision | Recall |
|---|---|---|---|---|---|
| ClinicDB | PraNet [42] | 91.22±0.42 | 88.29±0.43 | 94.26±0.33 | 91.59±0.32 |
| | Polyp-PVT [49] | 93.63±0.29 | 88.43±0.38 | 92.64±0.29 | 94.47±0.29 |
| | HarDNet-MSEG [29] | 93.05±0.36 | 88.34±0.42 | 92.72±0.35 | 93.16±0.33 |
| | ColonFormer-S [48] | 93.53±0.29 | 88.56±0.56 | 94.55±0.38 | 94.09±0.30 |
| | PVT-CASCADE [56] | 94.46±0.32 | 89.76±0.34 | 94.41±0.35 | 94.28±0.29 |
| | **Proposed MMCC-Net** | **94.55±0.21** | **90.24±0.29** | **94.75±0.26** | **94.69±0.22** |
| Kvasir | PraNet [42] | 90.42±0.26 | 85.51±0.36 | 89.52±0.39 | 92.09±0.36 |
| | Polyp-PVT [49] | 91.15±0.31 | 86.25±0.32 | 90.77±0.31 | 93.25±0.32 |
| | HarDNet-MSEG [29] | 91.28±0.27 | 85.59±0.28 | 92.68±0.30 | 92.29±0.29 |
| | ColonFormer-S [48] | 92.36±0.29 | 87.62±0.29 | 92.18±0.29 | 92.38±0.28 |
| | PVT-CASCADE [56] | 92.49±0.31 | 87.65±0.28 | 94.27±0.28 | 92.31±0.25 |
| | **Proposed MMCC-Net** | **92.29±0.22** | **88.35±0.26** | **94.75±0.25** | **92.81±0.24** |

For the ClinicDB dataset, MMCC-Net's accuracy improved by 17.42%, Dice score by 20.74%, and MIoU by 23.68% compared to U-Net. Compared to PraNet, MMCC-Net's accuracy increased by 1.96%, Dice score by 2.63%, and MIoU by 1.85%. MMCC-Net's performance over Polyp-PVT showed an improvement of 1.05% in accuracy, 0.81% in Dice score, and 0.20% in MIoU. Compared to HarDNet-MSEG, MMCC-Net's accuracy improved by 1.58%, Dice score by 1.16%, and MIoU by 1.82%. Compared to ColonFormer-S, MMCC-Net's accuracy was higher by 0.48%, Dice score by 1.10%, and MIoU by 0.72%. The improvements over PVT-CASCADE were 0.27% in accuracy, 0.04% in Dice score, and 0.15% in MIoU, while compared to FCB-Former, MMCC-Net's accuracy improved by 0.37%, Dice score by 0.06%, and MIoU by 0.19%. Compared to FCB-SwinV2, MMCC-Net's accuracy increased by 0.19%, Dice score by 0.02%, and MIoU by 0.08%. The results demonstrate the robustness and reliability of MMCC-Net, highlighting its superiority in terms of performance measures on the Kvasir and ClinicDB datasets.

Figure 7 illustrates the visual results of Experiment 2, highlighting that MMCC-Net has a notably lower rate of incorrect pixel predictions than other models. The analytical data for this experiment is provided in Table 3, demonstrating the models' stability in each evaluation metric. The results clearly show that MMCC-Net outperforms all other models regarding mDice, MIoU, precision, and recall across both datasets. MMCC-Net shows the highest consistency, achieving the lowest standard deviations in every evaluated metric on both datasets.

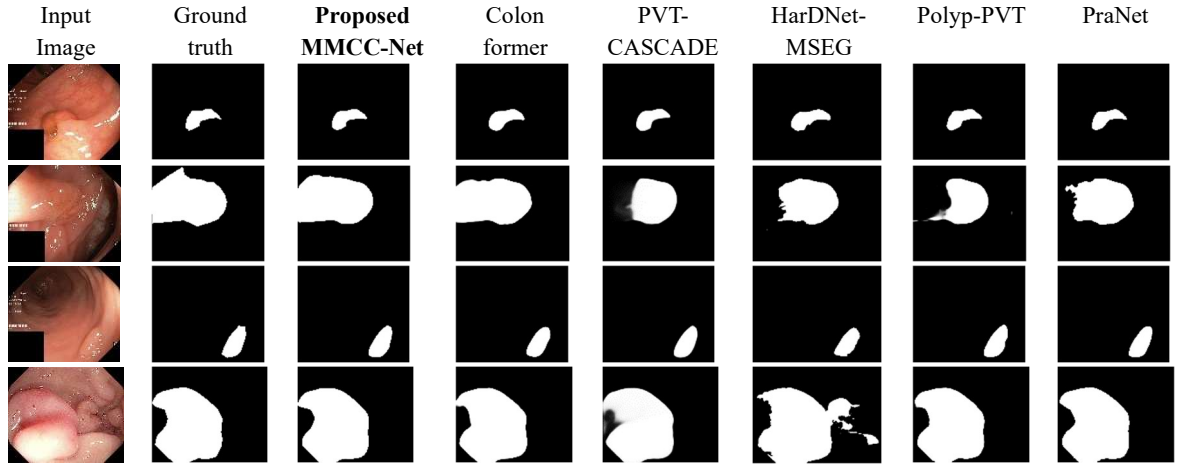

**Figure 7.** Comparative analysis of qualitative results from various models, based on the initial fold of the 5-fold cross-validation conducted on the Kvasir dataset.

### 4.3.2. Generalization capability

Our model's generalization capability was tested across four distinct datasets: CVC-ColonDB [60], CVC-300 [58], ETIS [59] and EndoCV2020 [57]. Figures 8 and 9 visually display the model's ability to generalize, using green to signify accurately identified polyps, yellow for correct polyp predictions, and red for incorrect predictions. Table 3 provides the quantitative analysis. Figure 8 shows the performance of the CVC-300 and ETIS datasets, with the first two columns presenting the input images and corresponding mask. Analysis of Figure 7 reveals challenges in detecting small polyps or those whose color closely matches the background. Traditional methods like UNet [65] and PraNet [42] fall short due to their lack of global context interaction and multi-scale feature analysis, leading to missed polyp detections or inaccuracies. ColonFormer [29] often misidentifies polyps, posing potential setbacks in clinical diagnosis. Despite improvements in segmentation accuracy from columns 3 to 10, issues persist, such as mistaking surrounding tissue for polyps in models like Polyp-PVT [48]. Our model employs a hierarchy of local features enhanced by attention module, skip paths, and feature enhancers to grasp the global context, alongside a cross-scale feature interaction approach and refined boundary delineation through depth-wise concatenation, resulting in superior segmentation.

Figure 9 displays the qualitative analysis for each method applied to the CVC-ColonDB [60] and EndoCV2020 [57] datasets, characterized by their complex backgrounds, blurriness, and low contrast. Despite UNet's implementation of skip connections to enhance feature integration, it often misinterprets surrounding tissues. HarDNet-MSEG [29] captures multi-scale features through varied receptive fields within a pyramid structure, and PraNet [42] utilizes attention mechanisms to focus on the target area. However, both methods overlook critical boundary details in less conspicuous regions, leading to segmentation with fragmented boundaries and incomplete shapes. On the other hand, ColonFormer [48], Polyp-PVT [49], FCB-Former [47], FCB-SwinV2 [52], and PVT-CASCADE [56], employing transformer modules delve into the semantic links between target areas and their boundaries, yielding morphologically coherent segmentation results. Nevertheless, these models are prone to significant segmentation inaccuracies by neglecting global context and multi-scale feature interactions.

We also compared the FCB-Former [27] and the FCB-SwinV2 [32], both of which employ a transformer parallel to CNN. Our results demonstrate that the proposed MMCC-Net outperforms the FCB-Former and the FCB-SwinV2 in terms of accuracy, Dice, MIou, Precision, and Recall. The reason is that training for both models was performed without using data augmentation techniques. Note that the performances of FCB-Former or FCB-SwinV2 degraded when compared with the results reported in [47] and [52]. This is because data augmentation techniques were adopted. The approaches such as ColonFormer [48], Polyp-PVT [49], FCB-Former [47], FCB-SwinV2 [52], and PVT-CASCADE [56] show improvement over UNet, HarDNet-MSEG, and PraNet [42], by leveraging transformer technology, boundary constraints, and multi-scale feature integration, our method demonstrates superior performance, especially in scenarios with smaller targets. This enhanced performance stems from integrating skip connections, feature enhancers, and attention modules, which collectively bolsters spatial and locational awareness, thereby sharpening the model's target localization and reducing recognition errors. The attention module enriches global context and edge delineation, augmenting scene comprehension and polyp differentiation from adjacent tissues.

Table 4 provides a comparison of various models for polyp segmentation across four datasets: CVC-300, ETIS, ColonDB, and EndoCV2020, detailing their performance in terms of accuracy, Dice, and MIoU with confidence intervals. The U-Net model exhibited moderate performance across all four databases. For the CVC-300 dataset, MMCC-Net achieved a Dice of 91.29% and an MIoU of 88.22%, improving upon U-Net by 36.05% in Dice score and 34.08% in MIoU. When compared to PraNet, MMCC-Net's Dice score improved by 3.95% and MIoU by 5.53%. Against Polyp-PVT, MMCC-Net showed an improvement of 2.47% in Dice score and 2.19% in MIoU. Compared to HarDNet-MSEG, MMCC-Net's Dice score improved by 2.76% and MIoU by 7.57%. When compared to ColonFormer-S, MMCC-Net's Dice score improved by 1.49% and MIoU by 4.86%. Compared to PVT-CASCADE, MMCC-Net's Dice score improved by 2.43% and MIoU by 0.82%. MMCC-Net also improved over FCB-Former by 1.28% in Dice and 0.62% in MIoU and over FCB-SwinV2 by 0.97% in Dice score and 0.02% in MIoU.

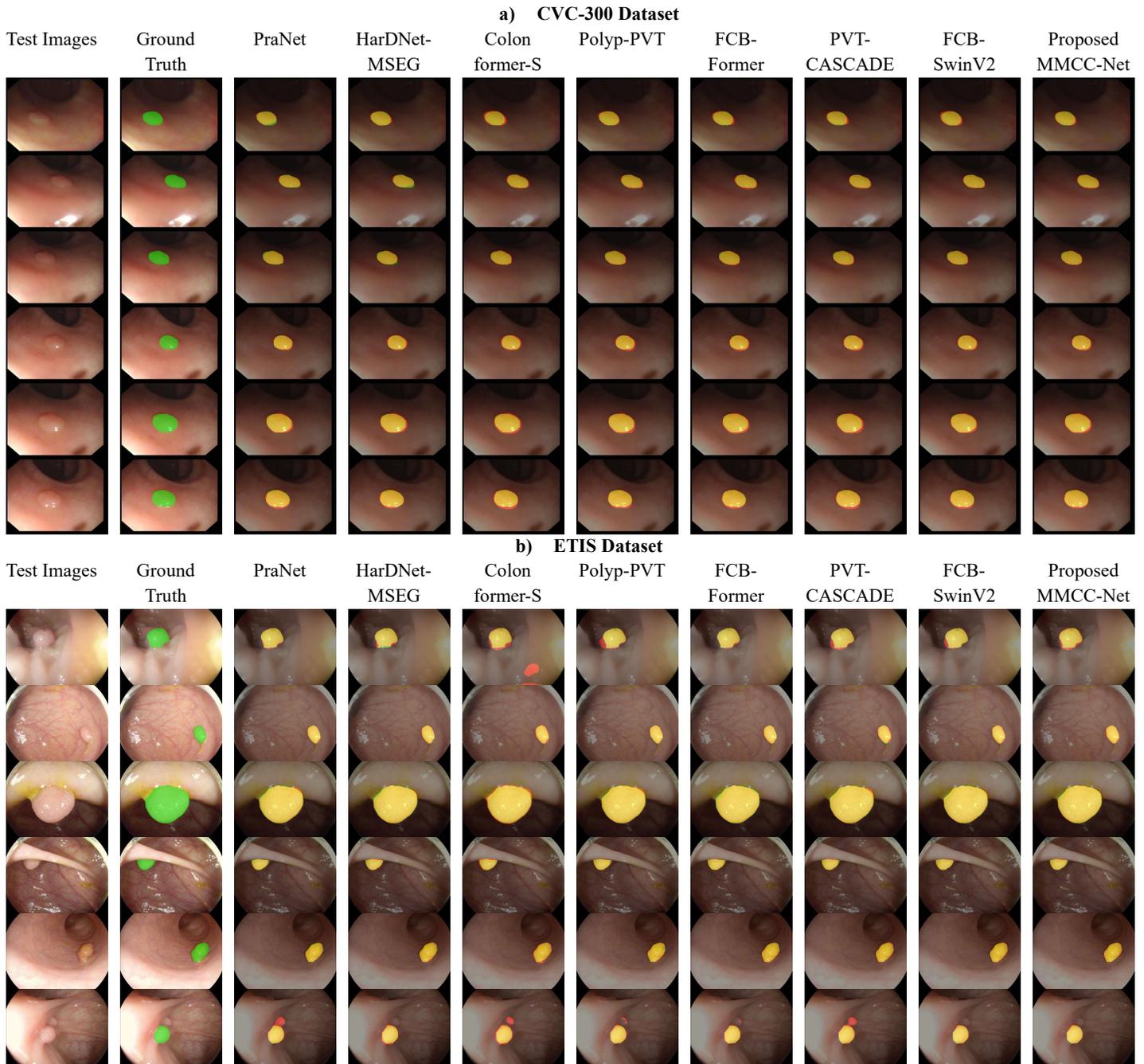

**Figure 8.** The comparisons of our proposed MMCC-Net and Seven SOTA models on the CVC-300 and ETIS datasets. Columns 1-2 show the original images and ground truth of the datasets; columns 3-9 show the segmented results of SOTA, and column 10 shows the results of the proposed MMCC-net. CVC-300 and ETIS datasets.

For the ETIS dataset, MMCC-Net achieved a Dice score of 80.38% and an MIoU of 72.71%, improving upon U-Net by 58.76% in the Dice score and 65.98% in MIoU. When compared to PraNet, MMCC-Net's Dice score improved by 12.27% and MIoU by 11.93%. Against Polyp-PVT, MMCC-Net showed an improvement of 1.95% in Dice score and 3.07% in MIoU.

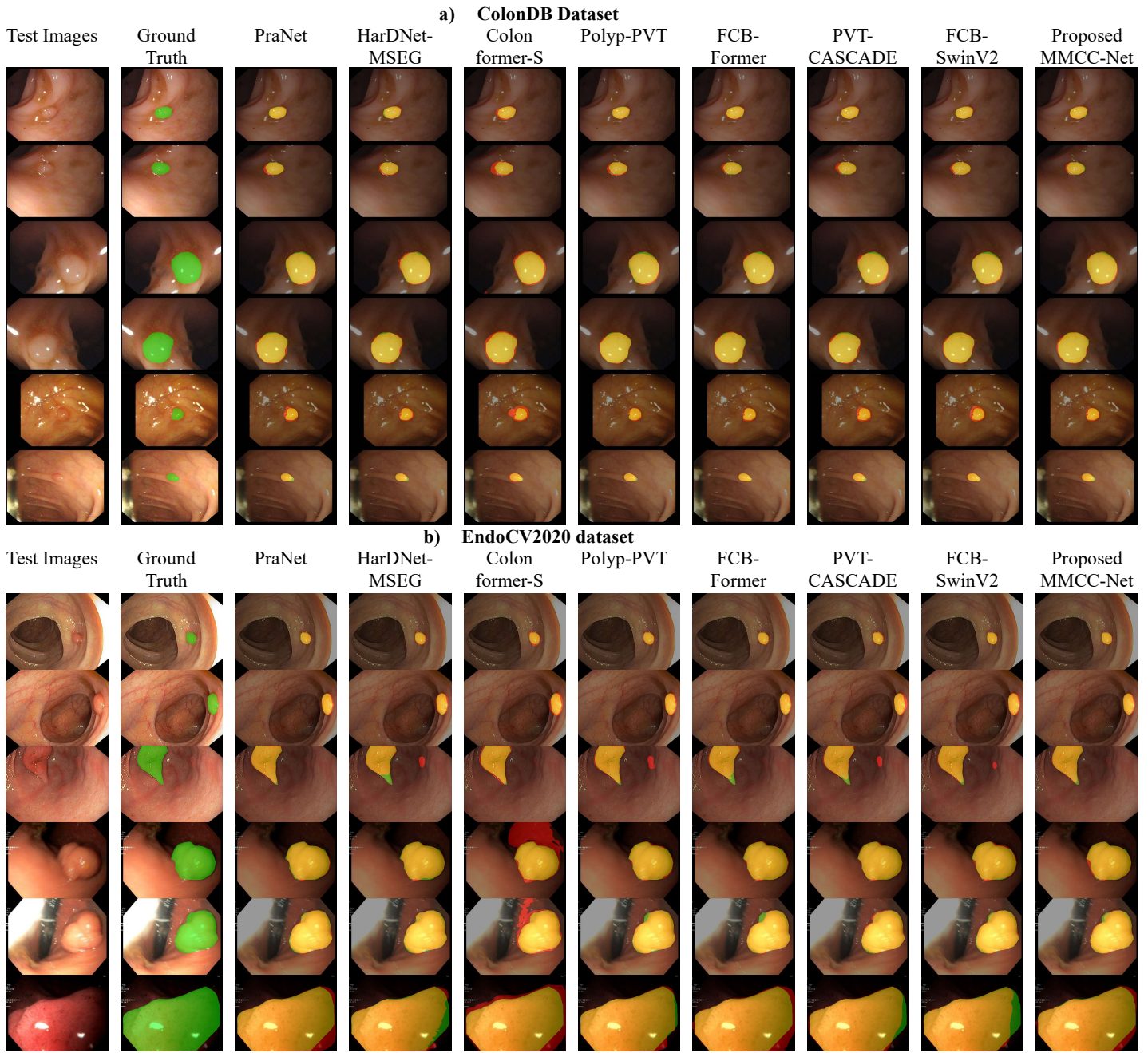

**Figure 9.** The comparisons of our proposed MMCC-Net and Seven SOTA models on the ColonDB and EndoCV2020 datasets. Columns 1-2 show the original images and ground truth of the datasets; columns 3-9 show the segmented results of SOTA, and column 10 shows the results of the proposed MMCC-net. CVC-300 and ETIS dataset datasets.

Compared to HarDNet-MSEG, MMCC-Net's Dice score improved by 18.83% and MIoU by 18.75%. When compared to ColonFormer-S, MMCC-Net's Dice score improved by 0.46% and MIoU by 0.50%. Compared to PVT-CASCADE, MMCC-Net's Dice score improved by 0.09% and MIoU by 0.15%. MMCC-Net also improved over FCB-Former by 0.06% in the Dice score and 0.14% in MIoU and over FCB-SwinV2 by 0.02% in the Dice score and 0.03% in MIoU.

Table 4. The comparisons of our proposed MMCC-net on Four publicly available datasets with SOTA for polyp segmentation with mean performance measures, standard deviation, and confidence interval (10 runs).

| Dataset | Method | Accuracy (Mean ± SD, 95% CI) | Dice (Mean ± SD, 95% CI) | MIoU (Mean ± SD, 95% CI) |
|---|---|---|---|---|
| CVC-300 [38] | U-Net [65] | 79.67 ± 0.45, (78.78, 80.78) | 67.13 ± 0.43, (66.12, 67.92) | 65.80 ± 0.42, (64.90, 66.50) |
| | PraNet [42] | 92.17 ± 0.21, (91.88, 92.68) | 87.82 ± 0.22, (87.21, 88.21) | 83.59 ± 0.31, (82.88, 84.08) |
| | Polyp-PVT [49] | 95.74 ± 0.15, (95.33, 95.93) | 89.09 ± 0.19, (88.84, 89.56) | 86.33 ± 0.22, (85.82, 86.62) |
| | HarDNet-MSEG [29] | 96.14 ± 0.14, (95.99, 96.51) | 88.84 ± 0.17, (88.43, 89.03) | 82.01 ± 0.20, (81.76, 82.48) |
| | ColonFormer-S [48] | 96.17 ± 0.15, (96.02, 96.54) | 89.95 ± 0.17, (89.76, 90.36) | 84.13 ± 0.19, (83.88, 84.60) |
| | PVT-CASCADE [56] | 97.29 ± 0.13, (96.99, 97.43) | 89.13 ± 0.15, (88.98, 89.50) | 87.50 ± 0.17, (87.31, 87.91) |
| | FCB-Former [47] | 97.32 ± 0.14, (96.89, 97.33) | 90.14 ± 0.16, (89.99, 90.51) | 87.68 ± 0.18, (87.49, 88.09) |
| | FCB-SwinV2 [52] | 97.08 ± 0.12, (96.97, 97.41) | 90.42 ± 0.14, (90.05, 90.57) | 88.20 ± 0.16, (87.79, 88.39) |
| | **Proposed MMCC-Net** | **97.36 ± 0.13, (97.03, 97.47)** | **91.29 ± 0.14, (91.16, 91.64)** | **88.22 ± 0.15, (87.86, 88.46)** |
| ETIS [39] | U-Net [65] | 76.85 ± 0.59, (75.76, 78.16) | 50.60 ± 0.52, (49.39, 51.59) | 43.79 ± 0.48, (42.68, 44.68) |
| | PraNet [42] | 92.56 ± 0.22, (92.05, 92.85) | 71.61 ± 0.26, (71.22, 72.22) | 64.96 ± 0.32, (64.25, 65.45) |
| | Polyp-PVT [49] | 96.13 ± 0.15, (95.94, 96.54) | 78.85 ± 0.20, (78.38, 79.10) | 70.54 ± 0.23, (70.25, 71.05) |
| | HarDNet-MSEG [29] | 93.40 ± 0.18, (92.93, 93.65) | 67.63 ± 0.21, (67.30, 68.18) | 61.23 ± 0.24, (60.84, 61.84) |
| | ColonFormer-S [48] | 94.35 ± 0.17, (93.94, 94.54) | 80.01 ± 0.18, (79.76, 80.48) | 72.35 ± 0.21, (71.84, 72.64) |
| | PVT-CASCADE [56] | 94.18 ± 0.15, (93.99, 94.59) | 80.31 ± 0.17, (79.91, 80.63) | 72.60 ± 0.22, (72.18, 72.98) |
| | FCB-Former [47] | 94.35 ± 0.16, (94.05, 94.65) | 80.33 ± 0.18, (79.97, 80.69) | 72.61 ± 0.19, (72.21, 72.86) |
| | FCB-SwinV2 [52] | 94.46 ± 0.14, (94.16, 94.76) | 80.36 ± 0.16, (80.01, 80.73) | 72.69 ± 0.21, (72.29, 72.89) |
| | **Proposed MMCC-Net** | **94.65 ± 0.15, (94.24, 94.84)** | **80.38 ± 0.17, (80.06, 80.78)** | **72.71 ± 0.19, (72.20, 73.00)** |
| ColonDB [40] | U-Net [65] | 72.78 ± 0.65, (71.59, 74.19) | 49.93 ± 0.56, (48.62, 51.02) | 43.62 ± 0.53, (42.41, 44.61) |
| | PraNet [42] | 90.32 ± 0.23, (89.81, 90.61) | 72.70 ± 0.24, (72.31, 73.31) | 64.83 ± 0.29, (64.12, 65.32) |
| | Polyp-PVT [49] | 93.13 ± 0.15, (92.94, 93.54) | 80.93 ± 0.17, (80.46, 81.18) | 72.82 ± 0.21, (72.31, 73.11) |
| | HarDNet-MSEG [29] | 92.36 ± 0.18, (91.89, 92.61) | 73.25 ± 0.22, (72.70, 73.58) | 66.13 ± 0.20, (65.52, 66.52) |
| | ColonFormer-S [48] | 95.17 ± 0.15, (95.04, 95.52) | 81.01 ± 0.17, (80.82, 81.42) | 73.11 ± 0.18, (72.86, 73.58) |
| | PVT-CASCADE [56] | 96.06 ± 0.14, (95.97, 96.37) | 82.45 ± 0.15, (82.28, 82.80) | 74.42 ± 0.16, (74.23, 74.83) |
| | FCB-Former [47] | 96.71 ± 0.15, (96.51, 96.91) | 82.49 ± 0.16, (82.33, 82.85) | 74.48 ± 0.17, (73.51, 74.91) |
| | FCB-SwinV2 [52] | 96.77 ± 0.12, (96.57, 96.97) | 82.51 ± 0.15, (82.35, 82.87) | 74.49 ± 0.16, (73.59, 74.99) |
| | **Proposed MMCC-Net** | **97.10 ± 0.11, (97.01, 97.41)** | **82.53 ± 0.14, (82.38, 82.90)** | **74.61 ± 0.13, (74.42, 75.02)** |
| EndoCV2020 [37] | U-Net [65] | 75.17 ± 0.57, (74.08, 76.48) | 67.31 ± 0.51, (66.10, 68.30) | 58.83 ± 0.52, (57.72, 59.72) |
| | PraNet [42] | 90.22 ± 0.23, (89.71, 90.51) | 74.15 ± 0.23, (73.76, 74.76) | 70.83 ± 0.27, (70.12, 71.32) |
| | Polyp-PVT [49] | 92.23 ± 0.16, (91.82, 92.42) | 76.30 ± 0.18, (76.05, 76.77) | 74.24 ± 0.22, (73.73, 74.53) |
| | HarDNet-MSEG [29] | 92.39 ± 0.15, (91.98, 92.58) | 75.01 ± 0.18, (74.76, 75.48) | 72.36 ± 0.21, (71.85, 72.65) |
| | ColonFormer-S [48] | 93.32 ± 0.12, (92.97, 93.45) | 76.13 ± 0.16, (75.94, 76.54) | 74.39 ± 0.17, (73.92, 74.64) |
| | PVT-CASCADE [56] | 94.00 ± 0.14, (93.91, 94.31) | 76.73 ± 0.14, (76.36, 76.88) | 74.38 ± 0.16, (74.19, 74.79) |
| | FCB-Former [47] | 94.15 ± 0.13, (93.95, 94.35) | 76.79 ± 0.15, (76.43, 76.95) | 74.52 ± 0.14, (74.22, 74.82) |
| | FCB-SwinV2 [52] | 94.31 ± 0.11, (94.11, 94.51) | 76.82 ± 0.16, (76.46, 76.98) | 74.59 ± 0.17, (74.29, 74.89) |
| | **Proposed MMCC-Net** | **94.84 ± 0.12, (94.75, 95.15)** | **77.43 ± 0.12, (77.08, 77.56)** | **75.39 ± 0.16, (74.98, 75.58)** |

For the ColonDB dataset, MMCC-Net achieved a Dice score of 82.53% and an MIoU of 74.61%, improving upon U-Net by 65.29% in the Dice score and 71.06% in MIoU. When compared to PraNet, MMCC-Net's Dice score improved by 13.50% and MIoU by 14.93%. Against Polyp-PVT, MMCC-Net showed an improvement of 2.60% in Dice score and 2.46% in MIoU. Compared to HarDNet-MSEG, MMCC-Net's Dice score improved by 9.44% and MIoU by 12.83%. When compared to ColonFormer-S, MMCC-Net's Dice score improved by 1.88% and MIoU by 1.76%. Compared to PVT-CASCADE, MMCC-Net's Dice score improved by 0.10% and MIoU by 0.19%. MMCC-Net also improved over FCB-Former by 0.05% in the Dice score and 0.17% in MIoU and over FCB-SwinV2 by 0.02% in the Dice score and 0.14% in MIoU.

For the EndoCV2020 dataset, MMCC-Net achieved a Dice score of 77.43% and an MIoU of 75.39%, improving upon U-Net by 15.12% in Dice score and 28.16% in MIoU. When compared to PraNet, MMCC-Net's Dice score improved by 4.42% and MIoU by 6.45%. Against Polyp-PVT, MMCC-Net showed an improvement of 1.48% in Dice score and 1.55% in MIoU. Compared to HarDNet-MSEG, MMCC-Net's Dice score improved by 3.24% and MIoU by 4.18%. When compared to ColonFormer-S, MMCC-Net's Dice score improved by 1.71% and MIoU by 1.35%. Compared to PVT-CASCADE, MMCC-Net's Dice score improved by 0.91% and MIoU by 1.01%. MMCC-Net also improved over FCB-Former by 0.84% in Dice score and 0.87% in MIoU and over FCB-SwinV2 by 0.80% in Dice score and 0.80% in MIoU. Our MMCC-Net consistently outperformed all other models and consistently achieved the highest performance across all datasets, confirming its robustness and reliability for polyp segmentation. The low SD and narrow CI indicate stable performance, making MMCC-Net a reliable choice for practical deployment in clinical settings.

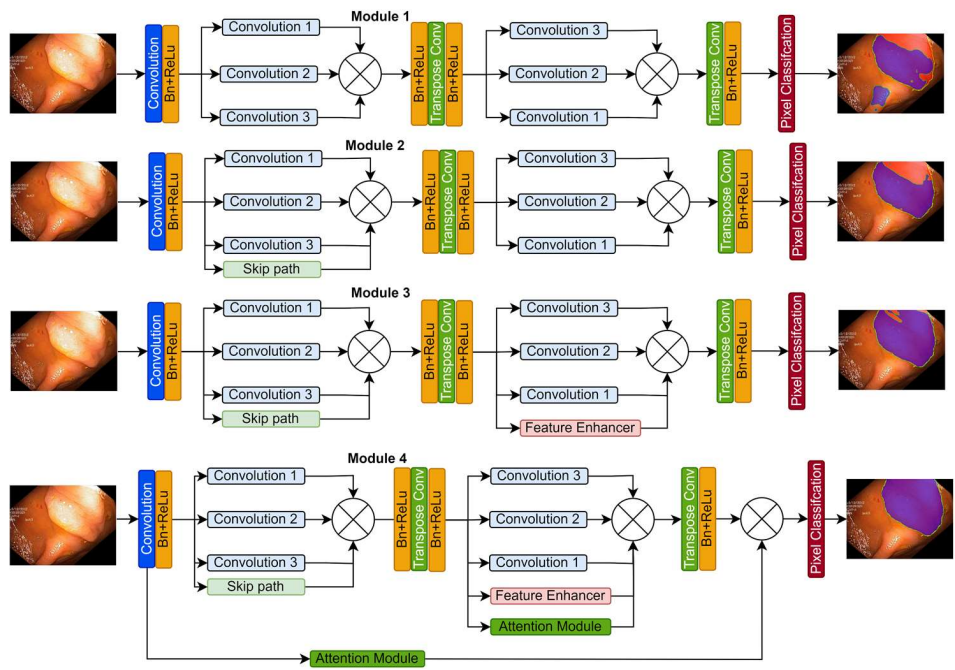

**Figure 10.** The MMCC-Net architecture with the integration of additional modules to handle challenging polyp cases.

### 4.3.3 Ablation Study of MMCC-Net

Experiments were conducted on the CVC-ClinicDB and Kvasir datasets to validate the performance of the MMCC-Net and the influence of attention modules, a feature enhancer, and a dense skip connection. The network modification performance on metrics such as Dice and MIoU across four distinct module configurations is reported in Table 5, and visual results are shown in $L_2$ loss on the Dice coefficient helps to provide a smooth gradient for optimization. This stabilizes the training process and ensures that the network is less likely to miss smaller but crucial segments of the minority class. We avoid the need to assign disparate weights to different classes explicitly. Instead, the Dice loss component inherently balances the influence of different classes based on their overlap, and the $L_2$ loss on the Dice coefficient further refines this balance. The loss functions, including cross-entropy loss, dice loss, joint loss, and joint loss with $L_2$ loss, were tested on the MMCC-Net. The corresponding results are given in Figure 11 and reported in Table 6.

Figure 10. We removed the dense skip routes, attention modules, and the feature enhancer in Network 1. Network 2 retains a single skip route, excluding attention modules and the feature enhancer. Network 3 integrates only one feature enhancer, and Network 4 is a more inclusive configuration with a single skip route, one feature enhancer, and two attention modules. A notable improvement in performance from Network 1 through to Network 4 highlights the substantial contribution of the dense skip connection, dual attention modules, and a feature enhancer to the system's efficacy.

**Table 5.** Performance outcomes of MMCC-Net on the Kvasir and CVC-ClinicDB databases with selective exclusions of feature enhancer, attention modules, and dense skip route.

| Network Types | Variations of Modules | CVC-ClinicDB | | Kvasir | |
|---|---|---|---|---|---|
| | | Dice | MIoU | Dice | MIoU |
| Network 1 | (Without attention module, skip path, and feature enhancer) | 85.28 | 83.92 | 81.72 | 77.22 |
| Network 2 | (With skip path) | 87.47 | 85.42 | 83.43 | 80.51 |
| Network 3 | (With feature enhancer) | 90.92 | 88.28 | 85.36 | 82.21 |
| **Network 4** | **(With skip path, both attention modules and feature enhancer)** | **94.41** | **90.11** | **92.55** | **88.42** |

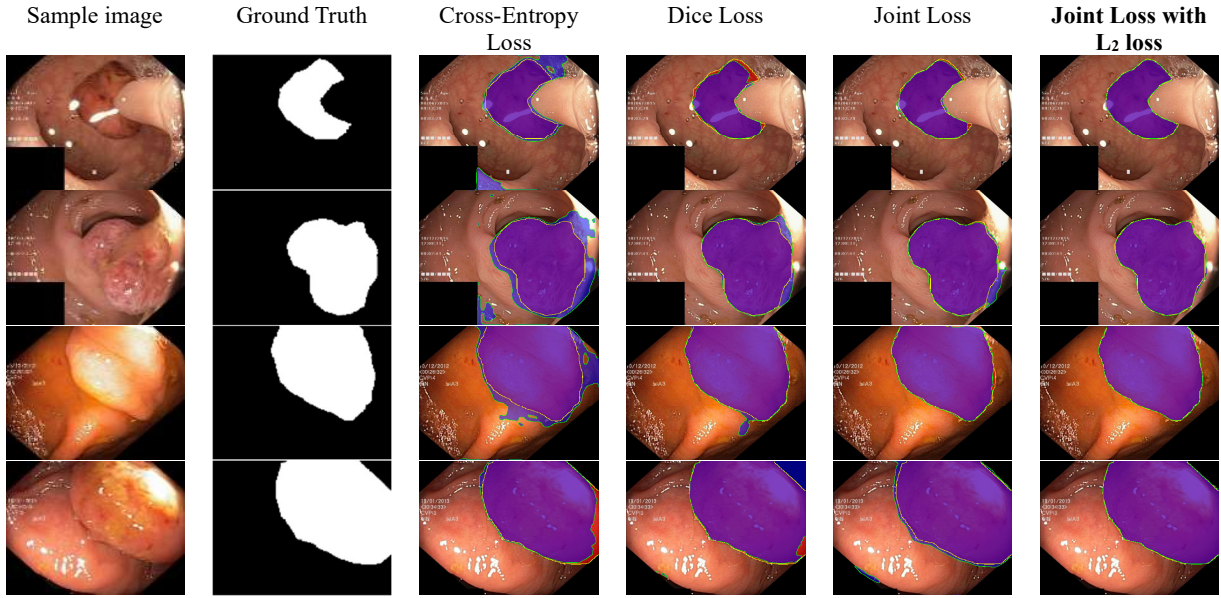

**Figure 11.** Sample images with corresponding masks demonstrate superior performance in accurately segmenting polyp boundaries, highlighting its effectiveness in handling class imbalance.

Notably, while dice loss exhibited superior performance over cross-entropy loss, the optimal results were observed with the joint loss approaches, especially with $L_2$ loss, highlighting its effectiveness in handling class imbalance. The effects of the SGD and Adam optimizers were also analyzed, and the results are shown in Table 7. The result reveals that the Adam optimizer marginally enhances segmentation performance. We also evaluated the performance of the MMCC-Net using various learning rates (LR). Table 8 shows that the learning rate $1e-4$ yields the best segmentation results regarding Dice, MIoU, and HDD. However, it's worth mentioning that our network critically depends on the learning rate.

Table 6. The effect of different loss functions on the performance of the MMCC-Net on the Kvasir and CVC-ClinicDB databases.

| Loss Function Types | CVC-ClinicDB | | | Kvasir | | |
|---|---|---|---|---|---|---|
| | Dice | MIoU | HDD | Dice | MIoU | HDD |
| Binary Cross-Entropy-loss | 91.78 | 87.81 | 16.98 | 86.12 | 85.14 | 19.74 |
| Dice loss | 93.48 | 88.74 | 9.40 | 89.71 | 86.21 | 16.21 |
| **Joint loss** | 93.81 | 89.12 | 5.71 | 91.32 | 87.26 | 8.28 |
| **Joint loss with $L_2$ Loss** | **94.41** | **90.11** | **4.07** | **92.40** | **88.28** | **6.11** |

Table 7. The segmentation performances of the MMCC-Net with Adam and SGD optimizers on the CVC-Clinic DB database.

| Optimizer Types | Dice | MIoU | HDD |
|---|---|---|---|
| SGD | 92.83 | 88.92 | 9.14 |
| **Adam** | **94.41** | **90.11** | **4.07** |

Table 8. The effect of different learning rates on the MMCC-Net performance on the CVC-ClinicDB database.

| Learning rate | Dice | MIoU | HDD |
|---|---|---|---|
| $3e-4$ | 94.22 | 88.43 | 14.85 |
| $4e-4$ | 93.48 | 88.24 | 12.91 |
| $5e-4$ | 93.41 | 89.98 | 11.25 |
| $2e-3$ | 92.52 | 87.45 | 10.82 |
| $1e-3$ | 93.78 | 89.81 | 10.45 |
| **$1e-4$** | **94.41** | **90.11** | **4.07** |

## 5. Discussion

Automated polyp segmentation in colonoscopy images has become crucial in ensuring precise polypectomy. Consequently, researchers and medical professionals have recently paid increasing attention to developing such methods. In our research, through an in-depth analysis of integrating attention mechanisms, skip connections, and feature enhancements, we identified two primary challenges: Firstly, neglecting intra-scale feature exploration may lead to overlooked detections among complex polyp images. Secondly, an absence of inter-scale feature interactions blocks the model's ability to learn from polyps exhibiting a broad spectrum of characteristics adaptively. Traditional approaches to polyp segmentation overlook the important link between cross-regional features, often resulting in segmentation that either misses or inaccurately defines boundaries. To tackle these limitations, the MMCC-Net, a novel framework for 2D polyp image segmentation, innovatively leverages multi-layer feature extraction and navigates them through a multi-pathway strategy, enabling the adaptive Insight of features. Our model is designed to address challenges like inconsistency in polyp sizes and shapes, offering a robust solution to enhance segmentation.

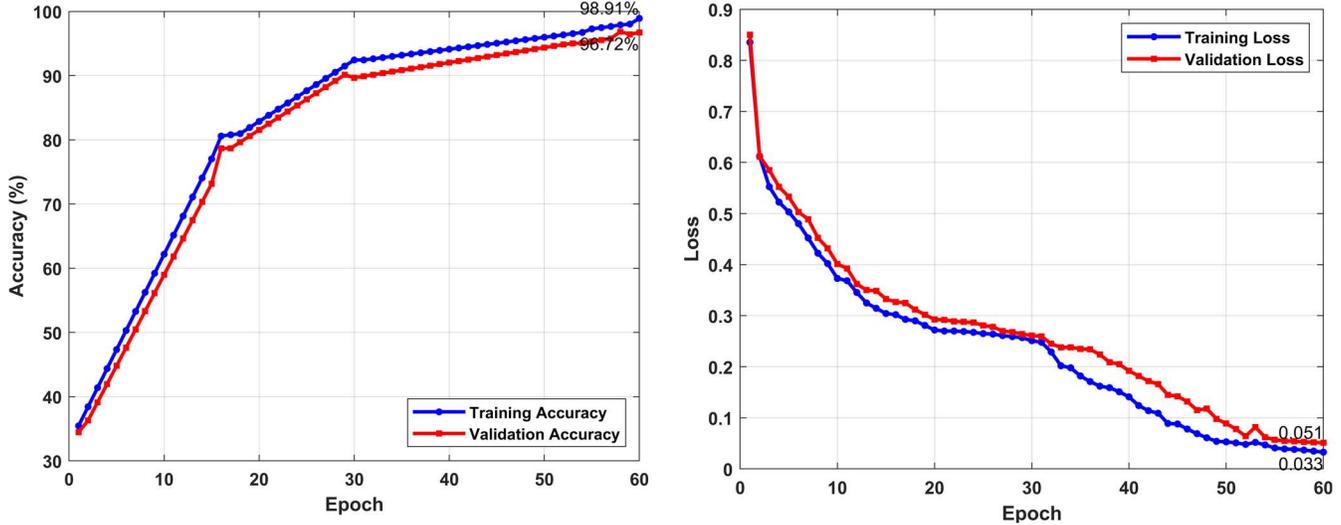

**Figure 12.** The training accuracy, validation accuracy, training loss, and validation loss of the proposed MMCC-Net with minimum and maximum values.

MMCC-Net distinguishes itself by employing features across three levels, a significant advancement over other methods that typically merge data from one or two adjacent layers. This approach enhances the network's capacity to accurately identify polyps of various sizes. A combination of cascade CNN, attention mechanisms, skip connections, and feature enhancements, local multi-scale features are meticulously extracted to bridge local and global feature relationships. Moreover, our model implements a novel three-scale feature fusion technique, facilitated by depth-wise concatenation, to encourage dynamic interaction among features at different stages of the network, thereby enriching the model's understanding of boundary and region correlations via dilated convolutions. The validity of our approach is underscored by comparative and ablation studies, confirming its efficacy. Our evaluation spanned six public polyp datasets, and both qualitative and quantitative analyses, including HDD and AUC metrics presented in Table 9 and training graphs illustrated in Figure 12, underscore our model's superior training and validation performance. MMCC-Net surpasses eight SOTA models in segmentation accuracy, paving the way for its potential clinical adoption as a reliable tool for polyp segmentation in colonoscopy images. The integration of MMCC-Net into clinical settings promises to significantly aid medical practitioners in their diagnostic processes.

The comparison of MMCC-Net against SOTA models in terms of Floating-Point Operations Per Second (FLOPs), the number of trainable parameters, training time, training accuracy for 60 epochs, Memory usage, inference time, and inference FLOPs were outlined in Table 10. Notably, MMCC-Net significantly reduces training time and the number of trainable parameters without a substantial increase in FLOPs, highlighting its efficiency and streamlined architecture.

Table 9. Quantitative comparisons between the MMCC-Net and eight SOTA models on six datasets in terms of Hausdorff distance (HDD) and area under the curve (AUC).

| Models | Kvasir | | CVC-ClinicDB | | CVC-300 | | ETIS | | ColonDB | | EndoCV2020 | |
|---|---|---|---|---|---|---|---|---|---|---|---|---|
| | HDD | AUC | HDD | AUC | HDD | AUC | HDD | AUC | HDD | AUC | HDD | AUC |
| U-Net [65] | 56.89 | 67.30 | 55.31 | 65.91 | 45.76 | 67.83 | 61.23 | 72.12 | 68.25 | 61.91 | 68.24 | 63.51 |
| PraNet [42] | 6.75 | 92.16 | 4.21 | 92.13 | 6.82 | 88.21 | 9.15 | 85.41 | 5.07 | 86.24 | 25.12 | 82.13 |
| Polyp-PVT [49] | 6.77 | 92.29 | 4.18 | 92.90 | 6.59 | 89.62 | 9.03 | 86.14 | 4.99 | 87.11 | 24.62 | 84.22 |
| HarDNet-MSEG [29] | 6.75 | 92.28 | 4.16 | 93.12 | 6.52 | 90.54 | 8.89 | 90.93 | 4.98 | 89.62 | 20.21 | 88.24 |
| ColonFormer-S [48] | 6.21 | 93.48 | 4.11 | 93.21 | 6.11 | 91.70 | 8.78 | 91.62 | 4.92 | 92.13 | 18.29 | 90.81 |
| PVT-CASCADE [56] | 6.15 | 93.59 | 4.10 | 93.80 | 5.56 | 93.21 | 8.55 | 92.11 | 4.90 | 93.24 | 16.48 | 92.62 |
| FCB-Former [47] | 6.14 | 93.76 | 4.09 | 93.89 | 5.53 | 93.49 | 8.53 | 92.48 | 4.91 | 93.61 | 15.97 | 92.92 |
| FCB-SwinV2 [52] | 6.12 | 93.79 | 4.08 | 93.96 | 5.51 | 93.61 | 8.51 | 93.62 | 4.90 | 93.88 | 15.92 | 93.21 |
| **Proposed MMCC-Net** | **6.11** | **93.81** | **4.07** | **94.21** | **5.49** | **93.90** | **8.49** | **94.13** | **4.89** | **94.42** | **15.81** | **93.41** |

Table 10. Comparison of Training time for 60 epochs, trainable parameters, inference time, and FLOPs across SOTA models.

| Models | FLOPs (G) | Parameters (Million) | Train Time (Hour) | Training Accuracy | Memory Usage (MB) | Inference FLOPs (G) | Inference Time (ms) |
|---|---|---|---|---|---|---|---|
| U-Net [65] | 75.97 | 31.04 | 11.0 | 87.82 | 100.1 | 45.58 | 21.2 |
| PraNet [42] | 13.15 | 30.49 | 7.0 | 95.21 | 124.5 | 7.89 | 23.0 |
| Polyp-PVT [49] | 10.00 | 25.10 | 10.5 | 96.27 | 97.8 | 6.49 | 22.2 |
| HarDNet-MSEG [29] | 11.38 | 33.34 | 8.6 | 95.18 | 166.3 | 6.83 | 24.3 |
| ColonFormer-S [48] | 16.03 | 33.04 | 10.0 | 96.89 | 379.4 | 9.62 | 26.8 |
| PVT-CASCADE [56] | 15.40 | 35.27 | 8.5 | 97.25 | 393.6 | 9.24 | 21.7 |
| FCB-Former [47] | 73.29 | 52.90 | 10.0 | 98.12 | 173.2 | 43.97 | 21.3 |
| FCB-SwinV2 [52] | 144.15 | 72.86 | 12.0 | 98.25 | 832.5 | 86.49 | 20.4 |
| **Proposed MMCC-Net** | **20.85** | **1.43** | **6.0** | **98.91** | **49.2** | **12.51** | **18.0** |

### 5.1 Practical Deployment of MMCC-Net in Real Clinical Settings

The MMCC-Net demonstrates significant promise in automated colorectal polyp segmentation, showing the potential to support gastroenterologists by enhancing the precision of Polyp removals. Figure 13 illustrates failure cases; it can be observed that MMCC-Net may produce incorrect predictions when the polyp is extremely large and appears similar to the surrounding normal tissues or with reflections or unclear boundaries. The publicly available datasets may not capture the clinical diversity and could limit the model's diagnostic applicability.

#### 5.1.1. Integration with Existing Medical Imaging Systems

Integrating MMCC-Net into medical imaging systems is crucial for practical deployment in hospitals. Compatible with standard Picture Archiving and Communication Systems (PACS) and Radiology Information Systems (RIS), MMCC-Net is designed to integrate seamlessly into existing workflows for retrieval, analysis, and storage of colon images using DICOM standards. Specifically, we are developing APIs enabling automated analysis of colonoscopy images in MMCC-Net through its integration with PACS/RIS. This will allow gastroenterologists to get access to segmentation results directly within their imaging systems, modify the workflow, and cut down the time spent on manual segmentation.

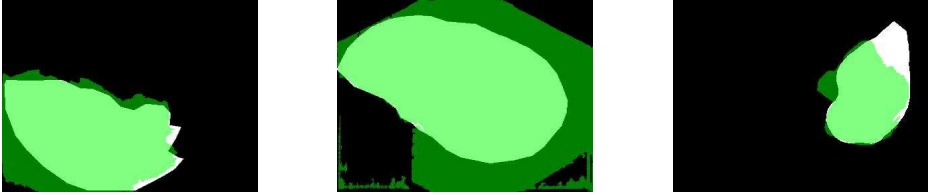

**Figure 13.** The failure examples of the proposed method are on Endocv2020 (left), Kvasir (middle), and CVC-ClinicDB (right) databases.

Based on potential user feedback, we intend to improve the generalization characteristics of our network by testing it on colonoscopy images from different systems. Apart from employing publicly available datasets like Kvasir, CVC-ClinicDB, ETIS-LaribPolypDB, CVC-300, EndoCV2020, and CVC-ColonDB, we also plan to collaborate with hospitals and research institutions to obtain diverse clinically relevant colonoscopy images. In this way, it can be determined whether MMCC-Net is robust or reliable enough under various imaging conditions and equipment; hence, it performs accurate polyp segmentation even for large-sized or reflecting unclear boundaries. Through this effort to solve these difficulties, we hope to reduce missed detections and false positives to improve the precision and efficiency of colorectal cancer screening and diagnosis.

### 5.1.2. Challenges of MMCC-Net in Clinical Processes

Compatibility issues can arise when integrating MMCC-Net with existing hospitals, PACS, or RIS. Achieving real-time image analysis of MMCC-Net requires computational resources and may cause hardware upgrades without compromising diagnostics speed or accuracy. Also, problems are faced when applying MMCC-Net in images from different hospitals due to issues of image resolution, size, and overall quality. The generalization of the MMCC-NET across diverse clinical environments is problematic given variations in imaging equipment and lighting conditions, as well as patient demographics, where model performance may be affected, making it difficult to segment polyps accurately across varied clinical settings. Moreover, some healthcare practitioners might shun new technologies, especially if they are already comfortable with traditional methods. To address these challenges, the model should contain robust preprocessing techniques and regularization strategies that enhance its adaptability and performance under various clinical contexts.

Comprehensive training sessions for gastroenterologists and clinical staff are necessary to ensure effective use of MMCC-Net. Initial training will be based on hands-on demonstrations, detailed tutorials, and user manuals to familiarize users with the system's functionalities. Regular training is equally important to update users with new features and practices through workshops and online courses. The detailed documentation that covers all aspects of MMCC-Net, from installation to troubleshooting, will provide users with a valuable resource for reference. Surveys and questionnaires can help collect user experiences related to usability, performance, and any issues encountered. This feedback provides useful insights into areas that may require improvement.

### 5.1.3. Analyze Common Failure Modes

We conducted a thorough error analysis to identify common failure modes to enhance the design of MMCC-Net. The primary failure modes include false positives, false negatives, and misclassified regions. This research has shown that the model fails in adequately segmenting minute polyps, large-sized ones, multiple ones, and irregularly shaped polyps, which are usually mistakes for surrounding tissue. However, if images are of poor lighting condition or low resolution or quality, then chances are high that those polyps will have been detected wrongly by the models. The model sometimes does not see small polyps, many of which possess irregular forms because they do not have sufficient information extraction, nor can they be separated from neighboring tissues easily. Such a problem may arise in case of wrong identification of their borders due to size differences when one tries to separate them from normal surrounding tissues.

Additional modules such as more sophisticated attention mechanisms, multi-scale feature integration, or preprocessing steps like histogram equalization, super-resolution algorithms, and noise reduction that normalize lighting conditions should help polish image resolution and enhance overall image quality, thereby improve feature detection for detecting polyps with higher accuracy. Small polyp extraction techniques combined with larger and irregularly shaped polyps can help to refine the MMCC-Net. More examples of small, huge, or irregularly shaped polyps in the training dataset and images taken under different lighting conditions and resolutions are needed to improve the model's robustness and generalization capabilities. Other regularization techniques against overfitting, such as dropout, data augmentation, and adversarial training, can also improve the model's generalization ability across different clinical settings.

In the future, we will train MMCC-Net on diverse medical images that would enable it to generalize better by use of images from various imaging equipment with divergent lighting conditions and wider population dynamics to ensure its functionality in multiple clinical settings becomes more reliable. An easy-to-use interface will be developed where MMCC-Net's outputs are integrated into the clinical workflow. MMCC-Net will be properly assessed after running extensive clinical trials and validations within real-world scenarios. Iterative improvements based on practitioner feedback will help ensure that MMCC-Net meets the practical needs of end-users.

## 6. Conclusion

MMCC-Net implements multi-scale features, two attention modules, and a feature enhancer (FE) to produce an effective multi-scale hierarchical feature representation. The FE combines with the attention modules and dense skip path segment smaller polyps in the images. The segmentation performance of MMCC-Net, which uses approximately 1.43 million trainable parameters and few filter-based convolutions, is competitive when implemented with joint loss by solving the class imbalance problem. MMCC-Net is far more effective than vision transformers since it doesn't require time-consuming pre- or post-processing of the images. Despite having far fewer trainable parameters than SOTA models, our tests show that MMCC-Net outperforms them on well-known benchmark datasets. The findings highlight MMCC-Net's exceptional performance with a confidence interval, with Dice scores ranging between 77.43 ± 0.12, (77.08, 77.56) and 94.45 ± 0.12, (94.19, 94.71) and Mean Intersection over Union (MIoU) scores ranging from 72.71 ± 0.19, (72.20, 73.00) to 90.16 ± 0.16, (89.69, 90.53), indicating a significant advancement over eight SOTA models. MMCC-Net achieved the highest Dice and MIoU scores on the CVC-ClinicDB dataset, with 94.41 and 90.11, respectively. MMCC-Net stands out not only for its efficiency but also for its reduced complexity, with its FLOPs amounting to approximately 20.85G. We have validated the effectiveness of the network modules, as evidenced by both quantitative and qualitative results.

**Declarations**
**Conflict of Interest:** The authors declared that they have no competing interests.